\documentclass[prb,aps,twocolumn,preprintnumbers,amsmath,amssymb,floatfix,longbibliography,superscriptaddress]{revtex4-2}
\usepackage{graphicx}
\usepackage{graphics}
\usepackage{psfrag}
\usepackage[pdfusetitle]{hyperref}
\usepackage{multirow}
\usepackage[sort&compress]{natbib}
\usepackage{amssymb}
\usepackage{amsmath}
\usepackage{amsthm}
\usepackage{bbold}
\usepackage{bbm}
\usepackage{enumerate}
\usepackage{textcomp}
\usepackage{orcidlink}

\usepackage[capitalize]{cleveref}

\usepackage{braket}
\usepackage[caption=false]{subfig}
\usepackage{siunitx}

\DeclareMathOperator{\Tr}{Tr}

\renewcommand{\i}{i}
\newcommand{\e}{e}
\renewcommand{\vec}[1]{\mathbf{#1}}
\newcommand{\id}{\mathbb{1}}

\newcommand{\tvec}[1]{\hat{T}_{\vec{R}_{#1}}}
\newcommand{\eik}[2][\vec{k}]{\e^{\i{#1}\cdot{#2}}}
\newcommand{\meik}[2][\vec{k}]{\e^{-\i{#1}\cdot{#2}}}

\newcommand{\we}{\omega_{\rm E}}
\newcommand{\hw}{\we}
\newcommand{\nmax}{N_{\rm max}}

\newcommand{\cre}[1]{\hat{c}^{\dagger}_{#1}}
\newcommand{\ann}[1]{\hat{c}^{\vphantom{\dagger}}_{#1}}
\newcommand{\creb}[1]{\hat{a}^{\dagger}_{#1}}
\newcommand{\annb}[1]{\hat{a}^{\vphantom{\dagger}}_{#1}}

\newcommand{\ketn}[2]{\ket{#1^{(#2)}}}

\newcommand{\psiblue}{\Psi_{\rm b}}
\newcommand{\psiorange}{\Psi_{\rm o}}
\newcommand{\psigreen}{\Psi_{\rm g}}
\newcommand{\psibrown}{\Psi_{\rm br}}
\newcommand{\psired}{\Psi_{\rm r}}

\newcommand{\alphb}{\alpha_{\rm b}}
\newcommand{\alpho}{\alpha_{\rm o}}
\newcommand{\alphg}{\alpha_{\rm g}}
\newcommand{\alphbr}{\alpha_{\rm br}}
\newcommand{\alphr}{\alpha_{\rm r}}

\usepackage[normalem]{ulem}   

\usepackage{xcolor}

\begin{document}

\hypersetup{pdftitle={Coherent-state ansatz for the Holstein polaron in one and two dimensions}}
\title{Coherent-state ansatz for the Holstein polaron in one and two dimensions}

\author{Connor M. Walsh\,\orcidlink{0009-0002-6311-9108}}
 \email[Contact author: ]{cmwalsh@ualberta.ca}
 \affiliation{Department of Physics, University of Alberta, Edmonton, Alberta, Canada}
 \affiliation{Theoretical Physics Institute, University of Alberta, Edmonton, Alberta, Canada}

\author{Igor Boettcher\,\orcidlink{0000-0002-1634-4022}}
 \email[Contact author: ]{iboettch@ualberta.ca}
 \affiliation{Department of Physics, University of Alberta, Edmonton, Alberta, Canada}
 \affiliation{Theoretical Physics Institute, University of Alberta, Edmonton, Alberta, Canada}
 \affiliation{Quantum Horizons Alberta, University of Alberta, Edmonton, Alberta, Canada}

\author{Frank Marsiglio\,\orcidlink{0000-0003-0842-8645}}
 \email[Corresponding author: ]{fm3@ualberta.ca}
 \affiliation{Department of Physics, University of Alberta, Edmonton, Alberta, Canada}
 \affiliation{Theoretical Physics Institute, University of Alberta, Edmonton, Alberta, Canada}
 \affiliation{Quantum Horizons Alberta, University of Alberta, Edmonton, Alberta, Canada}

\begin{abstract}
The Holstein model often serves as an archetype for electron-phonon interactions and polaron formation in solids. However, precise descriptions of the Holstein polaron are difficult when the phonon frequency is small and the electron-phonon coupling is strong, due to the presence of many phonons in the ground state. We present a semi-analytical approximation that consists of a variational ansatz with clouds of phonons surrounding the electron in the form of coherent states. This becomes particularly simple and exact in the Lang--Firsov limit. We determine the domain of validity away from this limit, and further explore the improvement achieved with a removal of the requirement that the phonon clouds form coherent states.
Both approximations work extremely well for the ground-state energy at strong coupling, and both work surprisingly well also at weak coupling. For the effective mass, however, both approximations have difficulty, particularly in the crossover region. Moreover, the coherent-state ansatz retains an unphysical jump in the crossover region. Nonetheless, the coherent-state ansatz provides a simple and intuitive picture of the polaron ground-state wavefunction, and in addition predicts accurate values for the ground-state energy.
\end{abstract}

\maketitle

\section{Introduction}\label{intro}

The formation of polarons, which are quasiparticles arising from electron-phonon interactions, has been a topic of much interest in condensed matter physics, originally due to its relevance for low-mobility semiconductors \cite{kuper62,langfirsov63} and later due to its application to the study of superconductivity \cite{alexandrov2007}. Perhaps the simplest model of polaron physics is the Holstein model \cite{holstein59,holstein59_2}, which treats the interaction between electrons and dispersionless optical phonons on a lattice. This model is well understood both near half filling and in the dilute limit, where it is simply known as the polaron problem. It is also most relevant for possible high-temperature superconductors, since it is high-frequency phonon modes that are expected to give rise to high critical temperatures and these typically arise in the form of optical modes, which are reasonably well modelled by the Holstein model.

In many respects the Holstein model is to electron-phonon interactions what the Hubbard model is to electron-electron interactions. Both focus on the (very) short-range part of the interaction and both are better suited to lattices than their long-range counterparts. Extensive reviews of the Holstein model can be found in Refs.~\cite{alexandrov_1995,fehske_2007,alexandrov_2010_chapter}. An excellent comparative summary of the various polaron models spanning the continuum limit (Fr\"ohlich \cite{frohlich_1950,frohlich1954}) to the lattice domain (Holstein \cite{holstein59,holstein59_2}) is provided in Ref.~\cite{franchini_2021}.

The first attempt at an exact solution for the Holstein polaron was by Ranninger and Thibblin in 1992 \cite{ranninger_1992}. They performed exact diagonalizations on a two-site model (see also Refs.~\cite{de_mello_97,marsiglio2022} for further studies of the spectral function and retardation effects in the same two-site model). This was followed by work on larger lattices \cite{marsiglio93,alexandrov_1994,marsiglio95,wellein_1996,capone_97,wellein_1997,fehske_1997,zhang_1999,fehske_2000,barisic_2004,barisic_2006}.

In 1999, Bon{\v c}a et al. \cite{trugman99} formulated an algorithm that used a controlled variational approach, essentially allowing a numerically exact solution for infinite lattices. This was expanded in a number of directions \cite{trugman_2001,trugman_Ku02,li_marsiglio_2010,li_2012,alvermann_2010,chen_2015,song_2015,song_2015_2,dunn_2019,chen_2019,kloss_2019,yan_2020,mitric_2023,gao_2025}, some of which will be explained further below. Many other approaches to approximately solve the Holstein problem can be found in Refs.~\cite{holstein_76,toyozawa_80,feinberg_1990,stephan_96,romero_99,alexandrov_2000,loos_2006,slezak_2006,berciu_2006,goodvin_2006,berciu_2007,barisic_2008,zoli_2010,berciu_2010,goodvin_2011,tayebi_2016,prodanovic_2019,jeckelmann_1998,hohenadler_2003,yang_2024,kornilovitch_1998,kornilovitch_1999,alexandrov_99,brown_97,romero_98,romero_99_mass,romero_2000,ciuchi_97,fratini_2001,fratini_2003,mitric_2022,zhao_97,kalosakas_1998,barisic_2002,de_filippis_2005,barisic_2007}. The various methods used in these publications will not be discussed further here.

In spite of the tremendous advances by Bon{\v c}a et al. \cite{trugman99}, it remains difficult to treat the strong-coupling regime, where a large number of phonons are needed to describe the ground state of the system, especially when the phonon oscillator frequency $\we$ is small. Importantly, this small-$\we$ regime is the domain where Migdal's approximation \cite{migdal_1958} is most valid. This approximation is usually invoked (as a theorem) to justify Eliashberg theory \cite{eliashberg_1960,eliashberg_1961}, which is believed to describe conventional superconducting materials \cite{allen_1983,marsiglio_2008}. It therefore makes sense that studies of the problem involving a single electron, which forms the building block for a many-electron system, should focus on this regime. Furthermore, the vast majority of Holstein polaron studies employ one-dimensional (1D) lattices, despite the fact that the 1D behaviour differs significantly from that in two and higher dimensions.

In this work, we treat the Holstein polaron by using an approximation scheme that evolves from the Lang--Firsov limit \cite{langfirsov63}. The Lang--Firsov limit (a name that we use interchangeably with the strong-coupling limit) is a ground-state wavefunction that is a Bloch superposition of localized coherent states, which we describe in detail later. Applying the Hamiltonian to the Lang--Firsov limit generates coherent states of phonons that neighbour the electron, and so we are motivated to explore an approximation consisting of families of these coherent states. In fact, if one simply examines the exact ground-state wavefunction, it immediately becomes evident that the components with the largest amplitudes occur in coherent-state-like families. 

In the following, we will explore two options, one in which actual coherent states are used (we call this the coherent-state ansatz (CSA)), and one in which the same Fock states comprising the coherent states are used, but with arbitrary coefficients to allow deviations from coherent states (we call this the restricted Hilbert space (RHS) approximation). These two variations, while both numerical, are extremely computationally efficient and as a consequence can be extended to higher-dimensional lattices at minimal computational cost. In both cases, we compare the principal features of the many-body wavefunctions and compute observables such as the ground-state energy or the polaron effective mass. This quantifies the validity of the various approximations as a function of coupling strength. For brevity, we constrain our discussion to one- and two-dimensional (2D) cubic lattices, aiming to highlight key differences between the two; it has been shown that the Holstein polaron in three dimensions exhibits qualitatively similar behaviour to its 2D counterpart \cite{trugman_Ku02,li_marsiglio_2010,alvermann_2010,li_2012}.

As a way to benchmark our approximative methods, we also present an algorithm for obtaining numerically exact results across almost all parameters in the model. The algorithm was introduced by Bon{\v c}a, Trugman, and Batisti{\'c} in \cite{trugman99} and refined in \cite{li_marsiglio_2010} to better address the strong-coupling regime.

The paper is structured as follows. We introduce the Holstein model for one electron in \cref{model}, establishing notation and relevant parameters. In \cref{methods}, we describe in detail the three methods, two approximative and one (numerically) exact. We then present in \cref{results} results for ground-state observables in 1D and 2D, highlighting important qualitative differences between these two cases. Specifically, we focus on the ground-state energy and effective mass, with additional discussion of the band structure and of the ground-state wavefunction. Our approximative methods excel in the strong-coupling regime but perform remarkably well for moderate and weak coupling strengths too, qualitatively predicting the nature of the crossover between weak- and strong-coupling regimes. We conclude with a summary in \cref{conclusion}.

\vspace{-8pt}
\section{Model}\label{model}

The Holstein model describes electrons on a lattice in the tight-binding formulation, with lattice ions modelled as independent Einstein oscillators of mass $M$ and frequency $\we$, and a linear coupling between electrons and lattice vibrations. The Hamiltonian reads
\begin{multline}
    \hat{H} = -t \sum_{\langle ij\rangle} \Bigl( \cre{i}\ann{j} + \cre{j}\ann{i} \Bigr) \\
    + \sum_j \biggl[ \frac{\hat{p}_j^2}{2M} + \frac{1}{2}M\we^2\hat{x}_j^2 \biggr] - \alpha \sum_j \hat{n}_j \hat{x}_j , \label{Hamiltonian}
\end{multline}
where $\cre{j}$ ($\ann{j}$) creates (annihilates) an electron at lattice site $j$, $\hat{n}_j = \cre{j}\ann{j}$ counts the number of electrons at site $j$, and $\hat{x}_j$ and $\hat{p}_j$ represent the position and momentum of the ion at site $j$. Here, $t$ is the hopping integral, setting the energy scale for the electron motion, and $\alpha$ denotes the strength of the electron-phonon coupling. In the first term, the sum is taken over all pairs of nearest-neighbouring sites, while the other sums are taken over all sites on the lattice. 

In what follows, we set Planck's constant $\hbar=1$. Using the standard relations
\begin{equation*}
    \hat{x}_j = \frac{1}{\sqrt{2M\we}} \bigl(\creb{j} + \annb{j} \bigr),\quad \hat{p}_j = \i\sqrt{\frac{M\we}{2}} \bigl(\creb{j} - \annb{j} \bigr),
\end{equation*}
the Hamiltonian (\ref{Hamiltonian}) can be re-written as
\begin{equation}
    \hat{H} = -t \sum_{\langle ij\rangle} \Bigl( \cre{i}\ann{j} + \cre{j}\ann{i} \Bigr) + \hw \sum_j \creb{j}\annb{j} - g\hw  \sum_j \hat{n}_j  \bigl(\creb{j} + \annb{j} \bigr) ,  \label{H_second_quant}
\end{equation}
where $\creb{j}$ ($\annb{j}$) creates (annihilates) a phonon excitation at site $j$ and we have introduced the dimensionless coupling constant $g={\alpha}/({\hw}{\sqrt{2M\we}})$.

While the parameter $g$ is useful in some settings, especially in strong-coupling perturbation theory, we mainly make use of the alternate dimensionless coupling constant $\lambda~=~{g^2\hw}/({2dt})$, where $d$ is the dimension of the lattice (in this work, $d=1,2$). The parameter $\lambda$ is used frequently in the literature, as it better characterizes the strong- and weak-coupling regimes (very roughly, $\lambda<1$ corresponds to weak coupling, $\lambda>1$ to strong coupling, and $\lambda\approx1$ to the crossover region). Throughout the paper, we measure energies in units of $t$, such that the system is parametrized by the two dimensionless quantities $\hw/t$ and $\lambda$.

For a lattice with $N$ sites, we specify the phonon states by an $N$-dimensional vector $\vec{n}$ of occupation numbers. Explicitly,
\begin{equation}
    \ket{\vec{n}} = \ket{n_1n_2\cdots n_N} = \prod_{j=1}^N \frac{(\creb{j})^{n_j}}{\sqrt{n_j!}}\ket{0} \label{position_basis}
\end{equation}
represents a configuration in which $n_1$ phonons occupy the first lattice site, $n_2$ phonons occupy the second site, and so on (the labelling of the $N$ sites is schematic, and thus can be applied to lattices in any dimension). The vacuum state without phonons is denoted by $\ket{0}$. The states $\cre{i}\ket{\vec{n}}$ form a complete orthonormal basis for the single-electron Hilbert space, with $i=1,2,\dots,N$. Even on a finite lattice, the complete Hilbert space is infinite-dimensional owing to the fact that the phonon numbers $n_j$ are unbounded. The number of phonons required to describe the ground state grows quickly with the coupling strength $\lambda$, especially for small frequencies $\we$. For example, the Lang--Firsov transformation \cite{langfirsov63} predicts that the average number of phonons is approximately $g^2=2d\lambda/(\hw/t)$.

The Hamiltonian (\ref{H_second_quant}) is translationally invariant and therefore can be block diagonalized in sectors with total momentum $\vec{k}$. Therefore, a convenient basis for the problem consists of the Bloch states
\begin{equation}
    \ket{\vec{k},\vec{n}} = \frac{1}{\sqrt{N}} \sum_{i=1}^N \eik{\vec{R}_i}   \hat{T}_{\vec{R}_i} \Bigl(\cre{1} \ket{\vec{n}} \Bigr) , \label{Bloch_state}
\end{equation}
where $\hat{T}_{\vec{R}_i}$ is the translation operator by the lattice vector $\vec{R}_i$. Throughout, we take site 1 to be the origin, such that $\vec{R}_{1}=0$. The states $\ket{\vec{k},\vec{n}}$ are constructed with periodic boundary conditions before taking the thermodynamic limit $N\rightarrow\infty$, whereupon the momentum $\vec{k}$ becomes a continuous variable taking on arbitrary values in the Brillouin zone.

\vspace{-8pt}
\section{Methods}\label{methods}

The basis states in \cref{Bloch_state} represent linear superpositions, where the state $\cre{1}\ket{\vec{n}}$ is copied onto all other lattice sites $i\neq1$ through translations. Here, $\cre{1}\ket{\vec{n}}$ is an object consisting of an electron at the origin and some surrounding phonons whose position relative to the electron is given by $\ket{\vec{n}}$. We emphasize that this choice of basis is distinct from the momentum-space representation obtained via Fourier transform; it is explicitly a real-space basis where phonons occupy definite positions with respect to the electron. To describe a given basis state, it is sufficient to specify the phonon configuration $\ket{\vec{n}}$, with the electron assumed to be at the origin. For a given value of $\vec{k}$, the Bloch state (\ref{Bloch_state}) can be uniquely re-constructed from knowledge of the vector $\vec{n}$.

For the methods employed in this work, this allows us to treat the infinite lattice directly, eliminating finite-size boundary effects. To do so, the infinite Hilbert space is truncated to exclude phonons on sites far from the electron, i.e. far from the origin. Furthermore, we impose an upper bound $\nmax$ on the number of phonons, discarding states with more than $\nmax$ phonons on any one site. This results in a finite-dimensional subset of the complete Hilbert space. The exact truncation procedure is different for each of the three methods, which we will describe shortly.

\begin{figure*}
    \centering
    \includegraphics[width=0.725\linewidth]{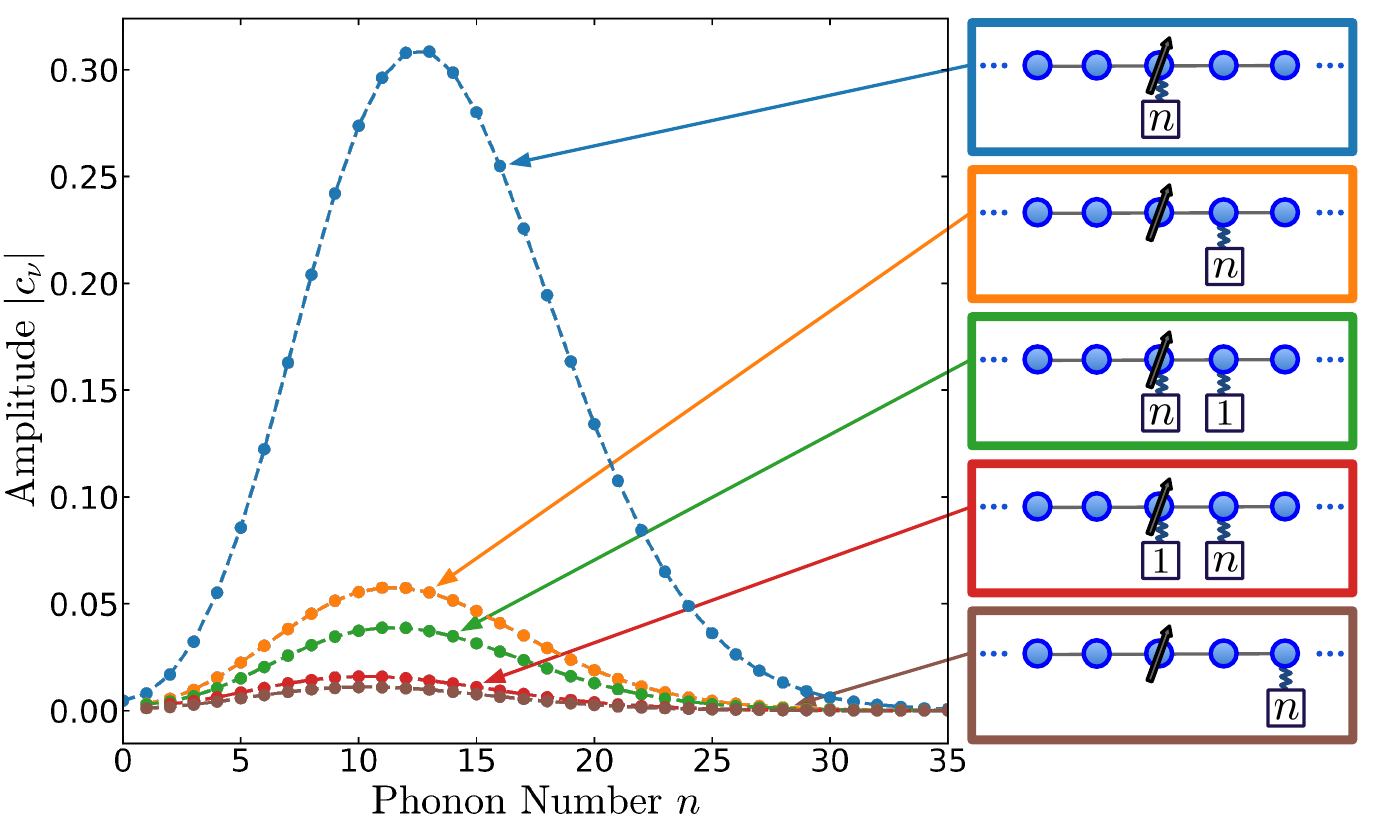}
    \caption{Plot of the numerically exact ground-state wavefunction for a 1D chain with $\hw=0.2t$ and $\lambda=1.5$, showing the amplitudes of states belonging to the five families depicted on the right. Dashed lines connect points within a given family and serve as a guide to the eye. States belonging to these families contribute the majority of the total probability density, with $\sum_{\nu} |c_{\nu}|^2=0.997$ for the five families depicted here. On the right, arrows represent the location of the electron, while the number of phonons on a given lattice site is indicated in the attached box, with $n=0,1,\dots$ labelling individual states within each family. Sites not explicitly labelled contain no phonons. As explained in the main text, the actual states used are Bloch superpositions, over all sites on the lattice, of the states depicted here. Except for those in the blue family (top), each state has an equivalent counterpart, obtained by reflecting across the electron site, which is also included in the appropriate family.}
    \label{fig_1D_states}
\end{figure*}

Our primary focus is on the ground state, which corresponds to $\vec{k}=0$. Examination of exact numerical results in the strong-coupling regime suggests that despite the presence of many phonons in the ground state, phonons tend to appear in highly specific configurations. As a result, only a relatively small subset of the Hilbert space is required to accurately describe the ground-state wavefunction, $\ket{\Psi_0}$. Specifically, the only basis states that are required consist of a large cloud of phonons occupying a single lattice site in the electron's vicinity, with exceptionally few phonons anywhere else in the system. Such states can be categorized into families, each parametrized by the number of phonons $n$ in the phonon cloud. The five families of states which contribute most prominently are depicted in \cref{fig_1D_states}, along with a plot of their corresponding amplitudes $c_{\nu}=\braket{\vec{k},\vec{n} | \Psi_0}$ for a particular choice of model parameters. Here the label $\nu=(\vec{k},\vec{n})$ is a generic index. Note that the results shown in \cref{fig_1D_states} are from the numerically exact result obtained with the modified form of the Bon{\v c}a--Trugman--Batisti{\'c} (BTB) method. These results show very systematic behaviour both in the set of families that contribute and in the smoothness of the $n$-dependence within a given family, but the individual coefficients in fact are allowed to vary freely to minimize the energy. Indeed, it is this observed smoothness that motivates the CSA described in the next section.

Throughout the paper, we will refer to these families by the colour of their surrounding boxes in \cref{fig_1D_states}. The hierarchy of families, from most to least important, can be pictured as follows (from top to bottom in the figure):
\begin{itemize}
    \item \textbf{Blue:} The electron and the phonon cloud occupy the same site, with no other phonons in the system.
    \item \textbf{Orange:} The electron and the phonon cloud occupy neighbouring sites, with no other phonons in the system.
    \item \textbf{Green:} The electron and the phonon cloud occupy the same site, with one additional phonon on a neighbouring site.
    \item \textbf{Red:} The electron and the phonon cloud occupy neighbouring sites, with one additional phonon on the same site as the electron.
    \item \textbf{Brown:} The electron and the phonon cloud occupy second-neighbouring sites, with no other phonons in the system.
\end{itemize}
The blue family of states is completely local, with all particles on the same site. States belonging to the other families, however, appear as inherently asymmetrical. For the sake of definiteness, the pictorial legend in \cref{fig_1D_states} depicts particles on sites to the right of the electron, but the equivalent states with particles on the left are also included in their respective families. Similarly, the 2D square lattice has four equivalent sets of states, obtained by $90^\circ$ rotations. Additionally, the brown family in 2D also includes states where the phonon cloud occupies one of the four states diagonally opposite the electron. Altogether, for a given value of $n$, the brown family then has eight members, one for each site that can be reached from the electron using two hops. We refer collectively to such sites as the electron's second neighbours.

\vspace{-12pt}
\subsubsection{Coherent-state ansatz (CSA)}\label{coherent}

In strong-coupling perturbation theory, the polaron ground state for $\lambda\to\infty$ is a coherent state given by the Lang--Firsov formula \cite{langfirsov63}
\begin{equation}
    \ket{\Psi_0}_{\rm LF} =
    e^{-\frac{1}{2}g^2}
    \frac{1}{\sqrt{N}} \sum_{i=1}^N  e^{g\creb{i}}\cre{i} \ket{0}\! . \label{LF}
\end{equation}
Using
\begin{equation}
\label{expansion}
    e^{g\creb{}}\ket{0} = \sum_{n=0}^\infty \frac{g^n}{\sqrt{n!}} \ket{n},
\end{equation}
where 
\begin{equation}
\label{fock}
    \ket{n} = \frac{(\creb{})^n}{\sqrt{n!}} \ket{0}
\end{equation}
are the normalized Fock states, it follows that a coherent state has amplitudes whose squares follow a Poisson distribution as a function of $n$, with mean and variance $g^2$.
A close examination of numerically exact ground-state wavefunctions in the strong-coupling regime (see \cref{sub_wavefunctions}), such as the one shown in \cref{fig_1D_states}, suggests that even for finite $\lambda$, the contributions from these blue states approximately resemble a coherent state, albeit with a modified coherence factor $\alpha<g$. Inspection of the exact results (see \cref{fig_1D_states}) indicates that other families also appear. In fact, using perturbation theory as described in Ref.~\cite{marsiglio95}, but now for the wave function, also generates these families, along with other states at each level of perturbation theory. This makes it difficult to identify these coherent states, and having the exact results certainly makes the identification easier.

Motivated by this observation, we propose the variational ansatz
\begin{equation}
    \ket{\Psi(\vec{k})} = w_{\rm b} \ket{\Psi_{\rm b}} + w_{\rm o} \ket{\Psi_{\rm o}} + w_{\rm g} \ket{\Psi_{\rm g}} + w_{\rm r} \ket{\Psi_{\rm r}} + w_{\rm br} \ket{\Psi_{\rm br}}\! , \label{ansatz}
\end{equation}
where $w_\mu$ are arbitrary coefficients and $\ket{\Psi_\mu}$ are coherent states corresponding to the families depicted in \cref{fig_1D_states}. The subscripts $\mu\in\{\rm b,\ o,\ g,\ r,\ br\}$ stand for the colours of the five families. In 1D, these coherent states for $\vec{k}=0$ are given by the expressions
\begin{align}
    \ket{\psiblue} &= \frac{B_{\rm b}}{\sqrt{N}} \sum_{i=1}^N \cre{i} e^{\alphb\creb{i}} \ket{0}\! , \label{psi_blue} \\
    \ket{\psiorange} &= \frac{B_{\rm o}}{\sqrt{N}} \sum_{i=1}^N \cre{i} \Bigl( e^{\alpho\creb{i+1}} + e^{\alpho\creb{i-1}}\Bigr) \ket{0}\! , \label{psi_orange} \\
    \ket{\psigreen} &= \frac{B_{\rm g}}{\sqrt{N}} \sum_{i=1}^N \cre{i} \Bigl( \creb{i+1} +\creb{i-1} \Bigr)e^{\alphg\creb{i}} \ket{0}\! , \label{psi_green} \\
    \ket{\psired} &= \frac{B_{\rm r}}{\sqrt{N}} \sum_{i=1}^N \cre{i} \creb{i} \Bigl( e^{\alphr\creb{i+1}} + e^{\alphr\creb{i-1}}\Bigr) \ket{0}\! , \label{psi_red} \\
    \ket{\psibrown} &= \frac{B_{\rm br}}{\sqrt{N}} \sum_{i=1}^N \cre{i} \Bigl( e^{\alphbr\creb{i+2}} + e^{\alphbr\creb{i-2}}\Bigr) \ket{0}\! , \label{psi_brown} 
\end{align}
where $\alpha_\mu$ are variational parameters and the $B_\mu$ are normalization factors.
In the above expressions, the two mirror-symmetric versions of each family are explicitly accounted for. To construct the CSA in 2D, the sums in \cref{psi_orange,psi_green,psi_red,psi_brown} are modified to include the four nearest neighbours and eight second neighbours as needed. We have written the CSA states here for $\vec{k}=0$, as appropriate for computing ground-state properties. When considering $\vec{k}\neq0$, the variational parameters $\alpha_\mu(\vec{k})$ and $w_\mu(\vec{k})$ explicitly depend on $\vec{k}$, and additional Bloch phases appear in \cref{psi_blue,psi_orange,psi_green,psi_red,psi_brown}. More details are provided in \cref{appendix_coherent}. 

To determine the ground-state energy, we vary the parameters $\alpha_\mu$ and the coefficients $w_\mu$ in Eq. (\ref{ansatz}) to minimize the energy expectation $\braket{\hat{H}} = \bra{\Psi}\hat{H}\ket{\Psi}$. We emphasize the simple form of the CSA wavefunction, which contains only nine free parameters (five variational parameters $\alpha_\mu$ and five weights $w_\mu$ with a normalization constraint) regardless of coupling strength, frequency, or dimension. In contrast, the RHS and BTB methods described below require a number of states on the order of thousands and millions, respectively. As we shall see in \cref{results}, the CSA yields ground-state energies which are remarkably similar to those from the RHS approximation, of which it is a special case. In addition to the fact that the computational complexity is independent of dimension or the choice of parameters, the CSA admits an intuitive picture of the polaron ground state: a superposition of coherent states whose positions relative to the electron are given by the legend in \cref{fig_1D_states}.
Moreover, while we have focussed on the Holstein model in this paper, the Lang--Firsov state and its generalization given by the CSA approximation are readily generalized to longer-range electron-phonon models such as the Fr{\"o}hlich model or other variations. For models where the coupling of the phonons is to the electron kinetic energy, it is presently unclear to us if this approach would succeed.

\vspace{-12pt}
\subsubsection{Restricted Hilbert space (RHS) approximation}\label{restricted}

The CSA described in the foregoing section imposes two main constraints, namely restricting the Hilbert space to just five families and constraining the individual families to take the form of coherent states. Maintaining the restriction to the same five families of states, we relax the second constraint, allowing the weight of individual states within a family to vary freely. We then compute the ground-state energy via exact diagonalization (ED): we refer to this as the RHS approximation.

As an example, for the blue family for $\vec{k}=0$ in the RHS approximation we write
\begin{align}
    \ket{\Psi_{\rm b}} = \frac{1}{\sqrt{N}} \sum_{i=1}^N \cre{i} \sum_{n\geq 0} d_n^{(\rm b)}\frac{(\creb{i})^n}{\sqrt{n!}}\ket{0}\!,
    \label{blue_example}
\end{align}
with arbitrary weights $d_n^{(\rm b)}$. The associated coherent state in the CSA has squared amplitudes that correspond to the Poisson distribution $|d_n^{(\rm b)}|^2 \propto (|\alphb|^2)^{n}/{n!}$, with a single variational parameter $\alphb$, which produces the term $e^{\alphb\creb{i}}\ket{0}$ in \cref{psi_blue} after summing over the phonon numbers $n$. Explicit expressions for all five families and arbitrary $\vec{k}$ are given in \cref{appendix_RHS}.

Restricting the Hilbert space to include only states from the five families results in a relatively small Hamiltonian matrix $H(\vec{k})$, which we diagonalize to find the lowest-energy eigenstate. The  necessary matrix elements $H_{\vec{m}\vec{n}}(\vec{k}) = \bra{\vec{k},\vec{m}} \hat{H} \ket{\vec{k},\vec{n}}$ are given by
\begin{multline}
    H_{\vec{m}\vec{n}}(\vec{k}) = -t\sum_{j} A_{1j} \eik{\vec{R}_{j}} \bra{\vec{m}} \hat{T}_{\vec{R}_{j}} \ket{\vec{n}}  \\
    + \delta_{\vec{m}\vec{n}} \hw \sum_j n_j -g\hw \bra{\vec{m}} (\creb{1}+\annb{1}) \ket{\vec{n}}\! , \label{H_mn}
\end{multline}
with $A$ the adjacency matrix of the lattice,
whose entries are defined as
\begin{equation}
    A_{ij} = \begin{cases}
        1 & \text{if sites $i$ and $j$ are nearest neighbours,} \\
        0 & \text{otherwise,}
    \end{cases} \label{A_mat}
\end{equation}
and with the sums taken over all lattice sites. 
The basis states $\ket{\vec{n}}$ are in the occupation number basis, as defined in \cref{position_basis}. The RHS is obtained by restricting the evaluation of the Hamiltonian matrix to include only those states $\ket{\vec{k},\vec{n}}$ belonging to the five families shown in \cref{fig_1D_states}. This corresponds to \cref{psib_RHS,psio_RHS,psig_RHS,psir_RHS,psibr_RHS} where the coefficients $\propto d_n$ of the states for each $n$ are kept as free variables. In practice, only values of $n$ less than a sufficiently large cutoff $\nmax$ need to be kept, which turns the Hamiltonian matrix into a matrix of size $\approx (4q+1)\nmax$ (i.e. {\bf not} $5 \times 5$, as would be the case for the CSA at this same level of approximation).

As a concrete example of these basis states, taken again from the blue family and for general ${\bf k}$, $(N_{\rm max}+1)$ of these states are given by
\begin{align}
    |\textbf{k},n\rangle = \frac{1}{\sqrt{N}} \sum_{i=1}^N \eik{\vec{R}_{i}} \cre{i}
    \frac{(\creb{i})^n}{\sqrt{n!}}\ket{0}\!,
    \label{blue_example_n}
\end{align}
and labeled with non-negative integers $n\leq N_{\rm max}$. The other $\sim4qN_{\rm max}$ basis states come from the other four families. Due to the relatively small size of the RHS, and to the linear scaling of the matrix dimension with $\nmax$, we can take $\nmax$ to be very large, ensuring properly converged results. Throughout, we ensure that all energies are converged to a relative error of less than $10^{-12}$.

The expression in \cref{H_mn} has a simple interpretation. Due to the matrix element $A_{1j}$, the first term sums over sites neighbouring the origin. The hopping term thus contributes a factor of $-t\eik{\vec{R}_{j}}$ if the phonon numbers $\vec{n}$ and $\vec{m}$ differ only by shifting all phonons by one site. The phonon term is diagonal and equal to $\hw N_{\rm ph}$, where $N_{\rm ph}$ represents the total number of phonons in the system. The coupling term contributes a factor of $-g\hw\sqrt{n_1+1}$ or $-g\hw\sqrt{n_1}$ if the states $\ket{\vec{n}}$ and $\ket{\vec{m}}$ differ by a single phonon at the origin but are otherwise identical. This last fact is due to the relations
\begin{equation}
    \creb{}\ket{n} = \sqrt{n+1} \ket{n+1}\!,\quad \annb{}\ket{n} = \sqrt{n} \ket{n-1}\!.
\end{equation}

To make the hopping term in \cref{H_mn} more concrete: the phase factor is $e^{\i k_x}$ if the two states $\ket{\vec{m}}$ and $\ket{\vec{n}}$ match after shifting the phonons in $\ket{\vec{n}}$ one site to the right; $e^{-\i k_x}$ if the states match after shifting one site to the left; $e^{\i k_y}$ for a shift of one site upward; and $e^{-\i k_y}$ for a shift of one site downward.

Of course, the number of families included in the RHS and the CSA is arbitrary. Greater accuracy can, in principle, be achieved by considering more than five families. However, as suggested by the plot in \cref{fig_1D_states}, the red and brown families already contribute relatively little. Indeed, the curves which correspond to the excluded families all lie well below the brown curve on the plot, and thus the error induced by their omission is small at strong coupling.

\vspace{-12pt}
\subsubsection{Modified Bon{\v c}a--Trugman--Batisti{\'c} (BTB) algorithm}\label{Trugman}

For numerically exact results against which to benchmark our approximation methods, we use the BTB algorithm \cite{trugman99} as modified in \cite{li_marsiglio_2010}. Here, a subset of the Hilbert space is generated by repeated application of the Hamiltonian to a given collection of states, known as the seed states. This then becomes the basis for a Hamiltonian matrix $H(\vec{k})$, which is used to determine the ground-state energy via ED.

The process of generating basis states is as follows. In the basis of Bloch states, the Hamiltonian has a diagonal piece, namely the phonon term $\hw\sum_j\creb{j}\annb{j}$, and an off-diagonal piece
\begin{equation}
    \hat{H}_{\rm off} = -t \sum_{\langle ij \rangle} \Bigl( \cre{i}\ann{j} + \cre{j}\ann{i} \Bigr) -g\hw \sum_j \cre{j}\ann{j} \bigl(\creb{j} + \annb{j} \bigr).
\end{equation}
We generate new states for our truncated basis by acting with $\hat{H}_{\rm off}$ on the existing states. Applying $\hat{H}_{\rm off}$ to a particular state yields:
\begin{itemize}
    \item a state with one additional phonon at the origin,
    \item a state with one fewer phonon at the origin, and
    \item $q$ states, where $q$ is the coordination number of the lattice, in which the electron has hopped from the origin to one of the neighbouring sites. 
\end{itemize}
Recalling that the basis states $\ket{\vec{k},\vec{n}}$ are superpositions of identical states $\cre{1}\ket{\vec{n}}$ copied onto all sites of the lattice, we choose a convention wherein we represent each basis state by $\cre{1}\ket{\vec{n}}$, its component with the electron at the origin. In order to maintain the appropriate projection, states resulting from a hop must be translated such that the electron is returned to the origin. In this way, the hopping portion of the Hamiltonian is effectively applied by shifting all phonons in the system by one lattice site.

This is perhaps most clearly seen through an example. Consider a state in 1D with 4 phonons on the site with the electron and 2 phonons on the site to its left, represented schematically by
\begin{equation}
    \ket{\cdots\ 0\ 2\ \hat{4}\ 0\ 0\ \cdots}\!, \label{seed}
\end{equation}
where the hat denotes the location of the electron. Note that $\hat{H}_{\rm off}$ consists of two pieces. The coupling piece has terms proportional to $\cre{j}\ann{j}\creb{j}$ and $\cre{j}\ann{j}\annb{j}$, which generate the states
\begin{equation}
    \ket{\cdots\ 0\ 2\ \hat{5}\ 0\ 0\ \cdots} \quad \text{and} \quad \ket{\cdots\ 0\ 2\ \hat{3}\ 0\ 0\ \cdots}\!, 
\end{equation}
respectively. The hopping piece yields the two states
\begin{equation}
    \ket{\cdots\ 0\ \hat{2}\ 4\ 0\ 0\ \cdots} \longrightarrow \ket{\cdots\ 0\ 0\ \hat{2}\ 4\ 0\ \cdots} \label{left_hop}
\end{equation}
and
\begin{equation}
    \ket{\cdots\ 0\ 2\ 4\ \hat{0}\ 0\ \cdots} \longrightarrow \ket{\cdots\ 2\ 4\ \hat{0}\ 0\ 0\ \cdots}\!, \label{right_hop}
\end{equation}
where we replaced the resultant states in \cref{left_hop,right_hop} by their equivalent counterparts with the electron at the origin. In this manner, the state in \cref{seed} generates the four states
\begin{equation}
\begin{aligned}
    \ket{\cdots\ 0\ 2\ \hat{5}\ 0\ 0\ \cdots}\!, \\
    \ket{\cdots\ 0\ 2\ \hat{3}\ 0\ 0\ \cdots}\!, \\
    \ket{\cdots\ 0\ 0\ \hat{2}\ 4\ 0\ \cdots}\!, \\
    \ket{\cdots\ 2\ 4\ \hat{0}\ 0\ 0\ \cdots}\!,
\end{aligned}
\end{equation}
where hops are represented heuristically by a shift of all phonons in the system. This procedure is easily generalized to a 2D lattice.

The process of building the truncated basis is iterative. We denote by $n_H$ the number of times we have applied the Hamiltonian, such that the seed states constitute the $n_H=0$ layer. Applying $\hat{H}_{\rm off}$ to each of the seed states generates a collection of candidate states for the next layer. From the candidates we remove any states which were already included in the seed, and the remaining states constitute the $n_H=1$ layer. If a state ever appears multiple times in the same layer, we keep only one copy.

The procedure is then carried out again, applying $\hat{H}_{\rm off}$ to all states in the $n_H=1$ layer and discarding any of the resulting states which belong to one of the previous layers. This yields the $n_H=2$ layer. The process is repeated until the desired number of layers $N_H$ is reached. The set of all states included in the layers $n_H=0,1,\dots,N_H$ becomes the basis for the Hamiltonian matrix $H(\vec{k})$, whose elements are computed according to \cref{H_mn}.

 The original work by Bon{\v c}a et al. used an electron on a bare lattice as the seed, represented schematically by
\begin{equation}
    \ket{\cdots\ 0\ 0\ \hat{0}\ 0\ 0\ \cdots}\!, \label{WC_seed}
\end{equation}
which yields extremely accurate results for small to moderate values of the coupling strength \cite{trugman99,trugman_Ku02}. However, this procedure breaks down at strong coupling, where the number of phonons needed to describe the ground state becomes very large. Motivated by the Lang--Firsov limit, we treat the strong-coupling regime by using the Lang--Firsov seed states
\begin{equation}
    \ket{\cdots\ 0\ 0\ \hat{n}\ 0\ 0\ \cdots}\!, \label{SC_seed}
\end{equation}
with $n=0,1,\dots,\nmax$ for a large but finite cutoff $\nmax$. These states are precisely the basis states needed to construct \cref{LF}. Though the notation is suggestive of a 1D lattice, these seeds work equally well in 2D and in higher dimensions. We note that the bare-lattice (or weak-coupling) seed is, in fact, contained in the Lang--Firsov (or strong-coupling) seed.

The size of the truncated basis grows linearly with the phonon truncation value $\nmax$ but exponentially with $N_H$, with the number of basis states given approximately by $(\nmax+1)\times(d+1)^{N_H}$. This demonstrates the importance of the strong-coupling seed. In 2D, for example, with $\hw=0.1t$ and $\lambda=1$, the number of phonons needed for converged results is approximately $90$. This limit is achieved efficiently through the strong-coupling seed, but to generate this many phonons from the weak-coupling seed would require $N_H\geq90$, resulting in a matrix dimension on the order of $3^{90}\approx 10^{43}$. Not only is this computationally infeasible, but it is also extremely wasteful: as we show in this work, very few of these states are actually required to achieve highly accurate strong-coupling results.

The BTB algorithm proves very efficient at generating appropriate basis states for describing the polaron ground state, by providing a natural tradeoff between the total number of phonons and their distance from the electron. Because the application of the Hamiltonian to a given layer can \emph{either} create more phonons \emph{or} move them away from the electron, states with the most phonons have these phonons on sites in the immediate vicinity of the electron, while states with phonons furthest from the electron contain a relatively small total number of phonons. 

Our most accurate results are obtained by using the bare-lattice seed for the weak-coupling regime and the Lang--Firsov seed for the strong-coupling regime. For intermediate coupling strengths, we employ both seeds and choose the one which yields the lower ground-state energy. Although the bare-lattice seed is contained in the Lang--Firsov seeds, the former allows us to include a larger $N_H$ in practice, and thus may give superior variational solutions. We ensure that results are converged in both the phonon truncation $\nmax$ and the number of layers $N_H$.

The computational time required to diagonalize a sparse matrix scales linearly with the matrix dimension \cite{koch_2015}. Increasing $N_H$ and $\nmax$ until results converge typically yields a Hilbert space of several million states and the resulting matrix can be diagonalized in times on the order of 30 seconds. This can become very time-consuming, particularly in the crossover region where convergence requires that both $N_H$ and $\nmax$ be large, or in more complicated calculations, for example with two electrons present, where the BTB strategy quickly runs into computational difficulty.
Since the size of the Hilbert space scales as $(d+1)^{N_H}$, increasing the number of layers even by one doubles (triples) the computation time in 1D (2D). Moreover, with typical required values of $N_H$ on the order of $10$ or $20$, scaling the problem to higher dimensions incurs enormous computational cost. This is another reason why the CSA could be very useful. Only a handful of variational parameters are required and computing the optimal parameters can be done in a fraction of a second. Furthermore, the computational time to compute observables with the CSA is constant, neither scaling with the parameters $\hw$ and $\lambda$ nor with the dimension of the lattice. The RHS exhibits an intermediate computational complexity, as the Hilbert space has a typical dimension of up to several thousand, and therefore the typical diagonalization time is roughly three orders of magnitude smaller. Here, the scaling with system dimension is linear (the number of states scales approximately as $(8d+1)\nmax$) and thus applying the RHS approximation to higher-dimensional systems is much more tractable than using the BTB approach.

\begin{figure*}[ht]
    \centering
    \subfloat[\label{fig_GS_energy_a}]{\includegraphics[width=0.50\textwidth]{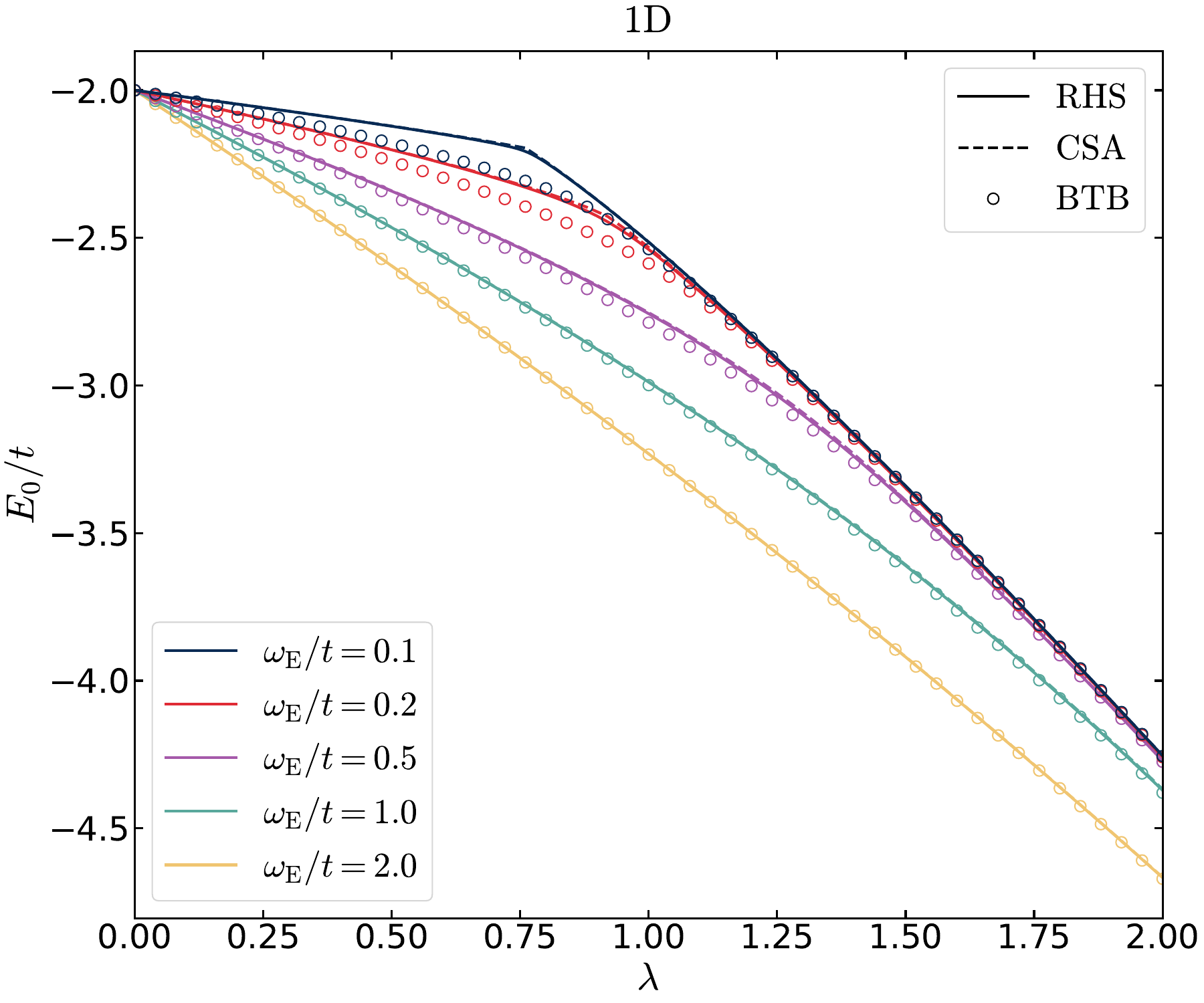}}
    \subfloat[\label{fig_GS_energy_b}]{\includegraphics[width=0.50\textwidth]{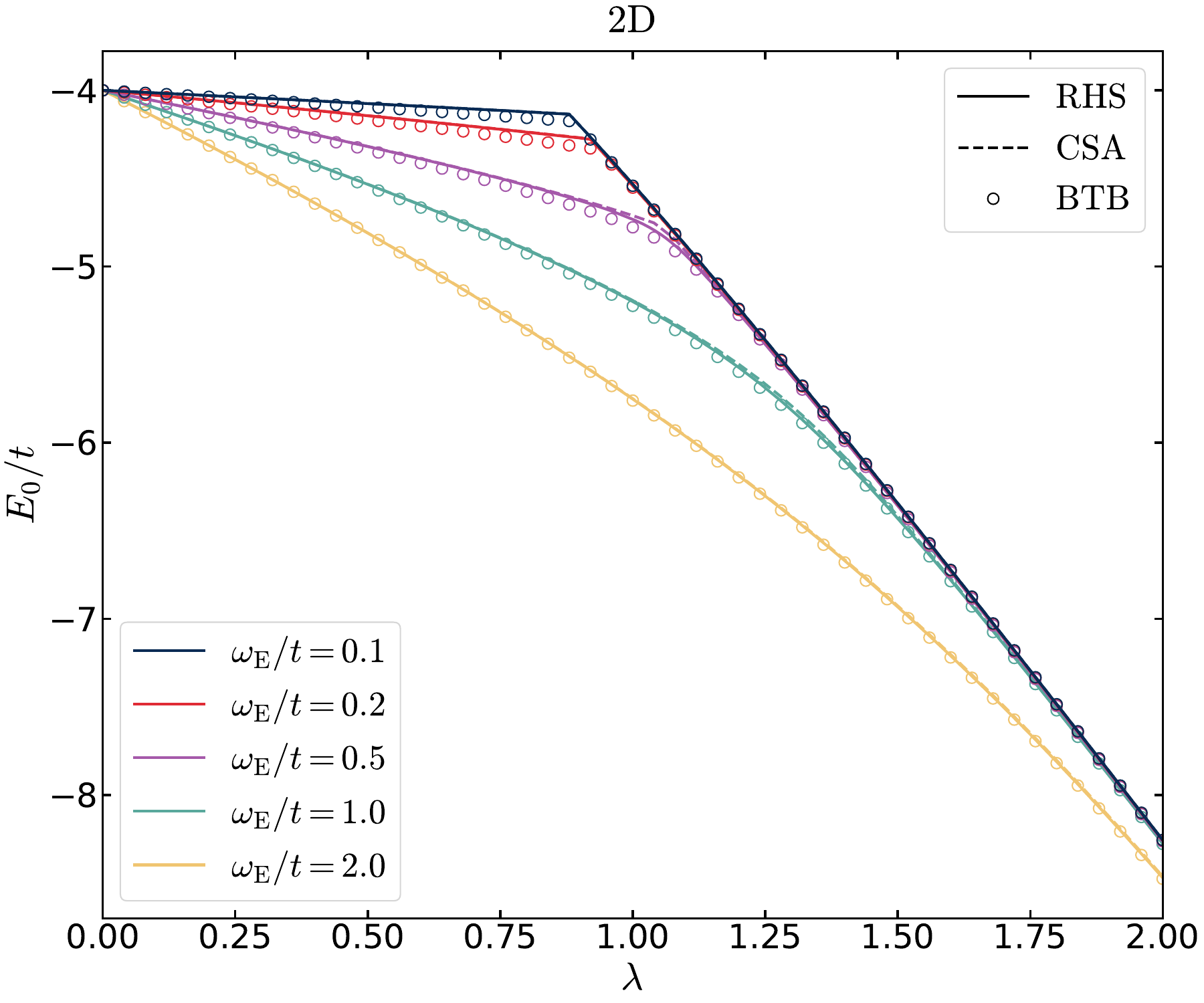}}
    \caption{Ground-state polaron energy $E_0$ as a function of coupling strength $\lambda$, in 1D (left) and 2D (right), for $\hw/t=0.1,\ 0.2,\ 0.5,\ 1.0,$ and $2.0$. Results from the CSA (dashed lines) and RHS (solid lines) are almost identical everywhere, and deviate most from the exact BTB results (circles) for low phonon frequencies and in the intermediate coupling regime. In 2D, results from both approximations remain very close to the exact results, in contrast to the 1D case. Note the apparent kink in the 2D energy curve for small frequencies. While the exact curves are analytic everywhere, the crossover between weak- and strong-coupling regimes is extremely abrupt, a fact which is captured well by both the CSA and the RHS.}
    \label{fig_GS_energy}
\end{figure*}

\vspace{-8pt}
\section{Observables}\label{results}

Here we present numerical results for the ground-state energy
\begin{equation}
    E_0 = \bra{\Psi_0} \hat{H} \ket{\Psi_0}
\end{equation}
and the effective mass $m^*$, which is defined by
\begin{equation}
    \frac{m}{m^*} \equiv \frac{1}{2t} \frac{\partial^2 E}{\partial k_x^2} \biggr\rvert_{\vec{k}=0} ,
\end{equation}
where $m$ is the bare electron mass. For the effective mass, we calculate the derivative via finite differences using small values of $\vec{k}$ (for the 2D lattice, we consider only the case where $k_y=0$).  We also show a few sample plots of the ground-state wavefunction $\ket{\Psi_0}$ and the energy dispersion $E(\vec{k})$, which is given by the smallest eigenvalue of the matrix $H(\vec{k})$. The ground-state energy corresponds to $\vec{k}=0$. The polaron energy and mass are known to be analytic functions of $\lambda$ for all $\we>0$ \cite{lowen_1988,lowen_gerlach_1988,lowen_1991}. In what follows, when we speak of abrupt changes in $E_0$ or $m^*$, it should be understood that they are smoothly varying quantities which nevertheless change rapidly over small $\lambda$-intervals.

\vspace{-12pt}
\subsubsection{Polaron energy and mass}

Plots of the ground-state polaron energy $E_0$ for a variety of $\we$-values are shown in \cref{fig_GS_energy}. The figure highlights a few key properties. As $\lambda\rightarrow0$, the ground state becomes a free electron on the lattice with energy $E_0=-2dt$. In the weak-coupling regime $\lambda\ll1$, the energy decreases linearly with $\lambda$, with a frequency-dependent slope. On the other hand, in the strong coupling limit $\lambda\gg1$ the energies become independent of $\we$ (the values of $\lambda$ shown are insufficiently large for the larger-frequency curves to converge, but they do in fact converge for sufficiently large $\lambda$).  Of particular interest is the crossover region $\lambda\lesssim1$, where a key difference between the 1D and 2D exact results (open circles in the plots) is highlighted: in 1D, the crossover is relatively gentle, without a clearly discernible point at which the strong-coupling regime begins. In 2D, by contrast, there is an abrupt crossover, especially for lower frequencies, which appears almost as a kink in the energy curve. While the deviation of the RHS energies (solid lines) from the exact values is largest in this crossover region, they nevertheless capture this behaviour qualitatively: a smooth crossover in 1D and an abrupt one in 2D. Across the whole parameter space, the CSA energy predictions (dashed lines) are nearly indistinguishable from the RHS ones.

\begin{figure*}[t]
    \centering  
    \subfloat[\label{fig_mass_logplot_a}]{\includegraphics[width=0.5\linewidth]{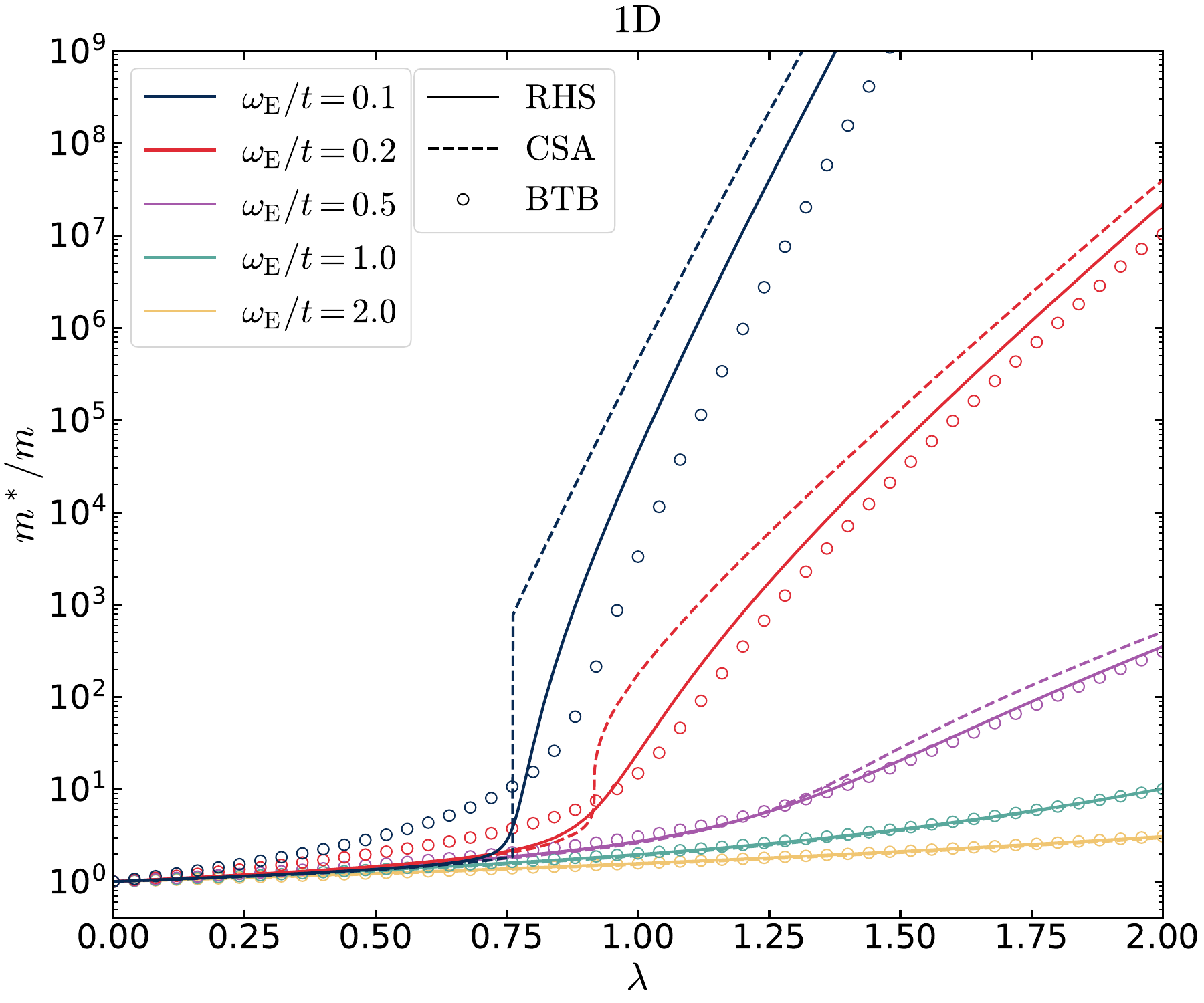}}
    \subfloat[\label{fig_mass_logplot_b}]{\includegraphics[width=0.5\linewidth]{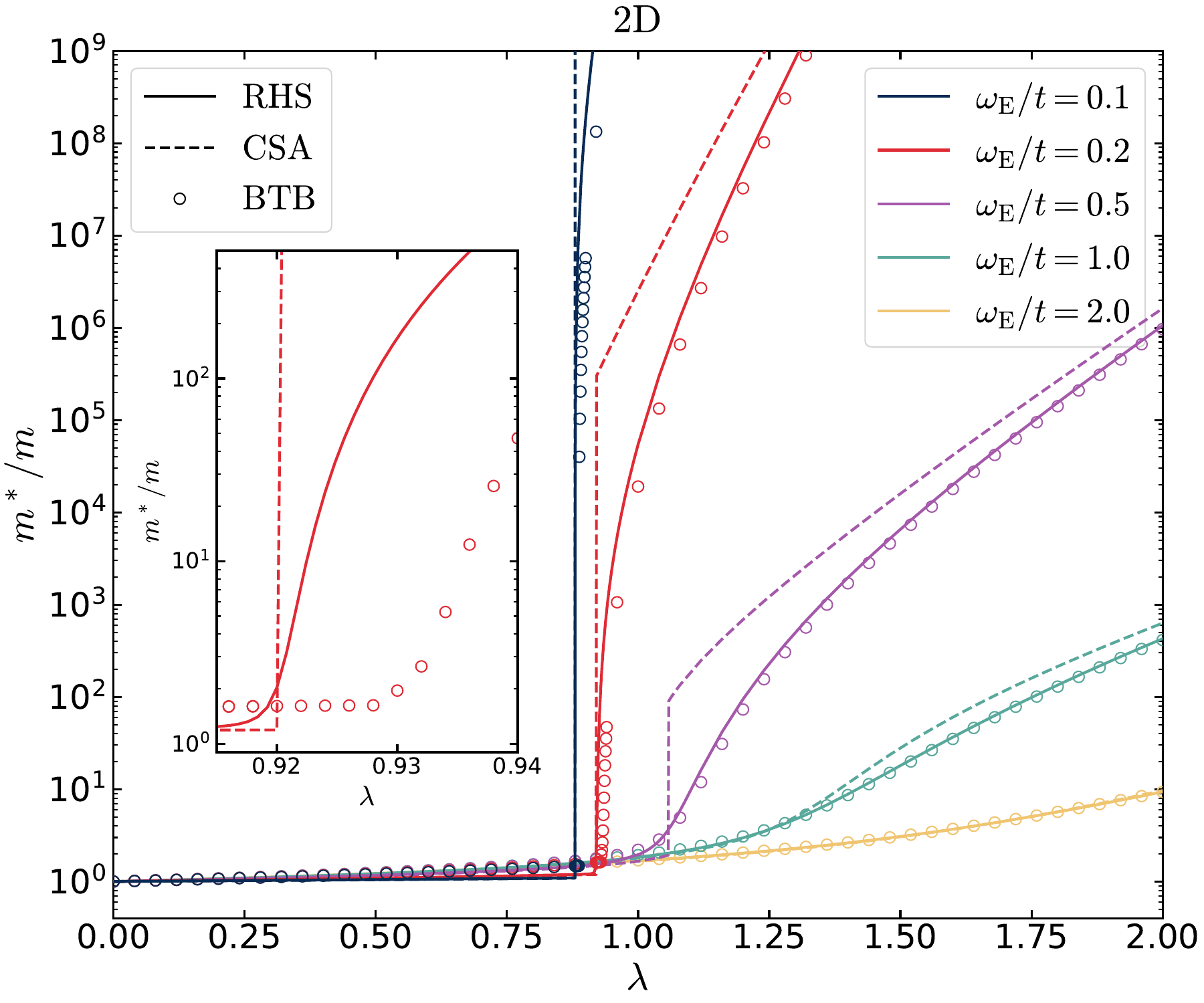}}  
    \caption{Effective mass $m^*$ as a function of coupling strength $\lambda$, in 1D (left) and 2D (right), for $\hw/t=0.1,\ 0.2,\ 0.5,\ 1.0,$ and $2.0$. Unlike the 1D case, the crossover from a weak- to a strong-coupling polaron in 2D is extremely abrupt, especially for small $\we$. Despite the abrupt crossover, the RHS mass prediction (solid lines) and exact BTB results (circles) vary smoothly with $\lambda$. This can be seen more clearly in the inset, where the crossover region for $\hw=0.2t$ is isolated. On the other hand, the CSA results (dashed lines) exhibit a discontinuous jump between weak- and strong-coupling regimes at a critical coupling strength $\lambda_{\rm c}$. The value of $\lambda_{\rm c}$ decreases with decreasing $\we$ in 1D, but tends to a finite value in 2D.}
    \label{fig_mass_logplot}
\end{figure*}

The polaron effective mass is plotted in \cref{fig_mass_logplot}. Most notably, the mass grows exponentially with $\lambda$, with more rapid growth for smaller values of the frequency $\we$. As was the case for the ground-state energy, the behaviour of the effective mass is markedly different in 1D and 2D. We first discuss the exact BTB results. In 1D, the mass increases smoothly with $\lambda$, having no definitive crossover point between weak- and strong-coupling regimes. In 2D, however, the mass remains small (on the order of the bare electron mass $m$) until a crossover value of $\lambda$ is reached, at which point the mass abruptly becomes very large. This is shown more clearly in the inset of \cref{fig_mass_logplot_b}, where the RHS and BTB curves increase from small to large mass over a narrow range of $\lambda$-values. The RHS results underestimate the mass in the weak-coupling regime and overestimate it in the strong-coupling regime, with the effect becoming more pronounced with decreasing $\we$. Still, the RHS captures the qualitative features discussed here, and performs quantitatively rather well, especially in 2D. To make the behaviour in the weak-coupling regime more apparent, we present alternate mass plots in \cref{appendix_mass} with linear rather than logarithmic scaling on the vertical axis.

For small values of $\we$, the effective mass predicted by the CSA differs considerably from that given by the exact solutions. The effective mass is clearly a more subtle indicator of accuracy compared with the ground-state energy and the CSA does a relatively poor job of replicating the BTB results except at large values of $\hw$ or very large values of $\lambda$ (not shown). In particular, in the crossover region, the CSA exhibits a jump discontinuity at a critical coupling strength $\lambda_{\rm c}$. On the other hand, constraining the solution to take the form of a coherent state turns out to work well for the ground-state energy in both the weak- and strong-coupling regimes.
Near the crossover, though, the exact ground state interpolates smoothly between the two regimes. While the RHS is able to mirror this interpolation, the CSA is forced to choose either a weak- or a strong-coupling polaron, and thus no smooth interpolation is possible. In spite of its shortcomings, the CSA nonetheless accurately reproduces the wavefunction at strong coupling, as will be shown below.

\vspace{-12pt}
\subsubsection{Ground-state wavefunctions}
\label{sub_wavefunctions}
\begin{figure}[ht]
    \centering
    \includegraphics[width=1.0\linewidth]{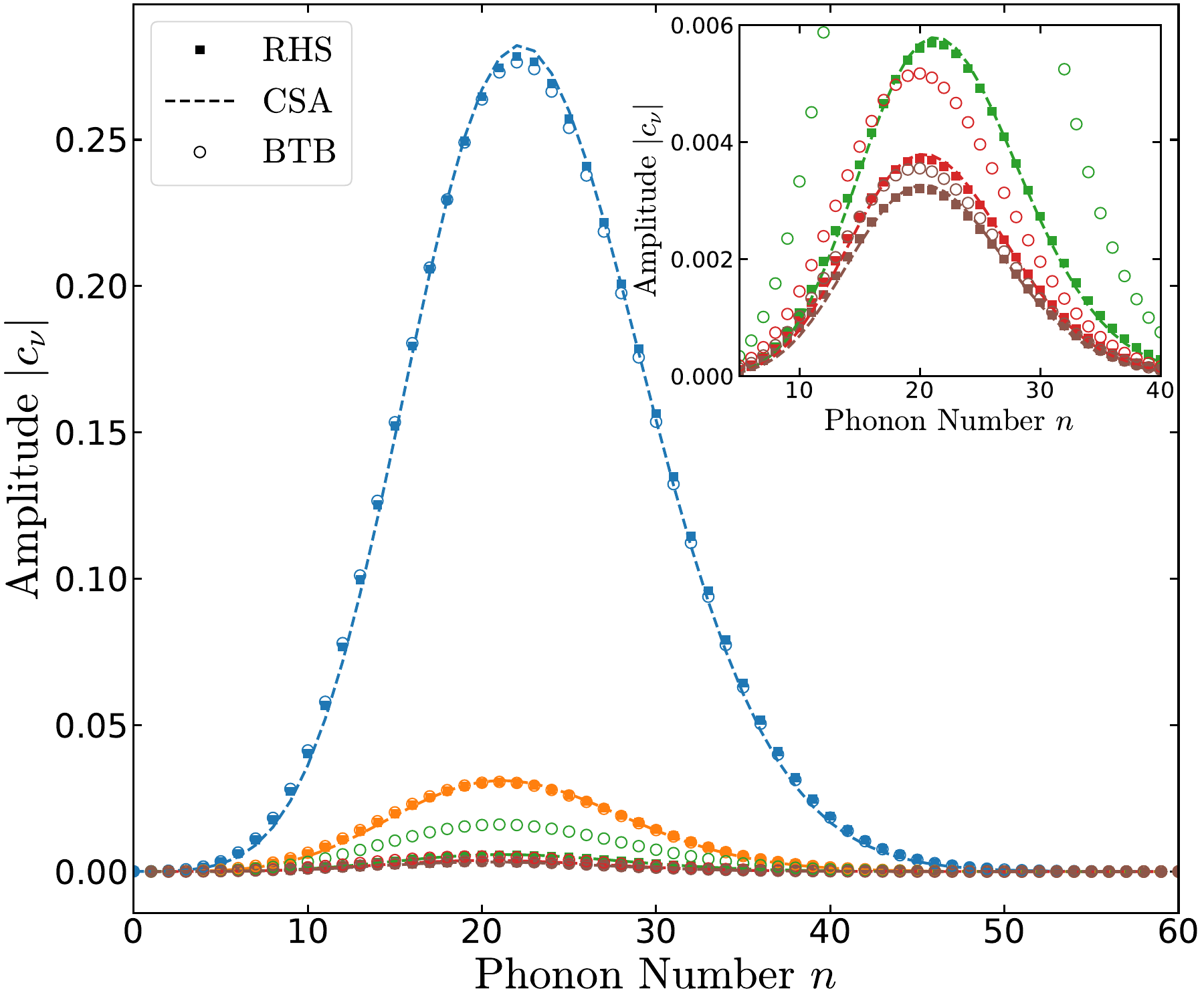}
    \caption{Plot of the largest components of the ground-state wavefunction for $\hw=0.2t$ in 2D, at strong coupling (here $\lambda=1.25$). The colour of the lines and symbols indicates which of the families from \cref{fig_1D_states} is represented. For the BTB wavefunction (open circles), we only show amplitudes for the states belonging to the five families included in the RHS (filled squares). These families contribute $\sum_\nu |c_{\nu}|^2=0.9989$ of the total BTB probability, with the largest amplitude among the states not included here being $0.0019$. Results from the CSA are shown with a dashed line for clarity, but the amplitude is defined only at integer values of $n$. The blue and orange curves agree well for all three methods. The inset shows the lowest-lying curves in more detail. While these do not match precisely, the overall shape of the coherent states, as well as the relative importance of the families, are consistent across all methods. This demonstrates the power of the RHS approximation.}
    \label{fig_GS_psi_2D}
\end{figure}

While much of our discussion has focused on the accuracy of our approximation methods, we emphasize that the wavefunctions generated by these methods give rise to a simple, intuitive picture of the phonon configurations found in the ground state. To highlight this, we show a sample wavefunction in \cref{fig_GS_psi_2D}. The plot shows the 2D result for $\hw=0.2t$ and $\lambda=1.25$, though wavefunctions across the whole strong-coupling regime (as well as in 1D) are qualitatively similar. Evidently, both approximate methods faithfully capture the two most important families. Though agreement for the other three families is less precise, the overall shape of the coherent states, as well as the relative heights of the curves, are apparent with all three methods.

One particularly attractive feature of the CSA is that it is exact in both limits $\lambda\rightarrow0$ and $\lambda\rightarrow\infty$. In the weak-coupling limit, the ground state is a free electron on a bare lattice, which corresponds to $\ket{\psiblue}$ with $\alphb=0$. On the other hand, the strong-coupling limit gives rise to the Lang--Firsov state \cref{LF}, which is again described by $\ket{\psiblue}$, this time with $\alphb=g$. The insight suggested by our results is that in between these two limits other families emerge, also taking the form of coherent states. Despite being designed specifically to target the strong-coupling regime, our variational solution also describes the weak-coupling regime reasonably well, in both one and two dimensions.

\begin{figure*}[t]
    \centering
    \subfloat[\label{fig_GS_psi_2D_several_lam_a}]{\includegraphics[width=0.32\textwidth]{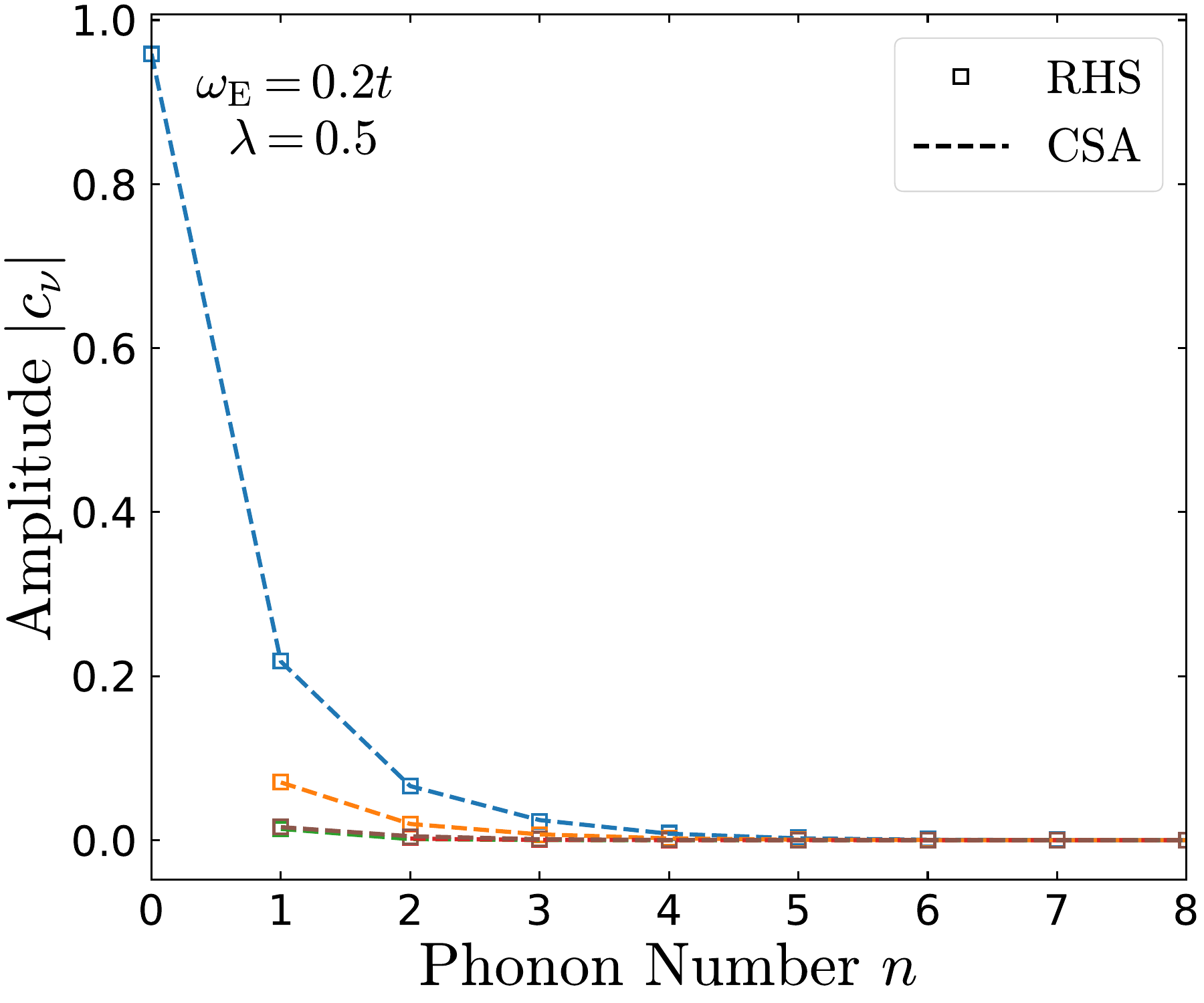}}
    \subfloat[\label{fig_GS_psi_2D_several_lam_b}]{\includegraphics[width=0.32\textwidth]{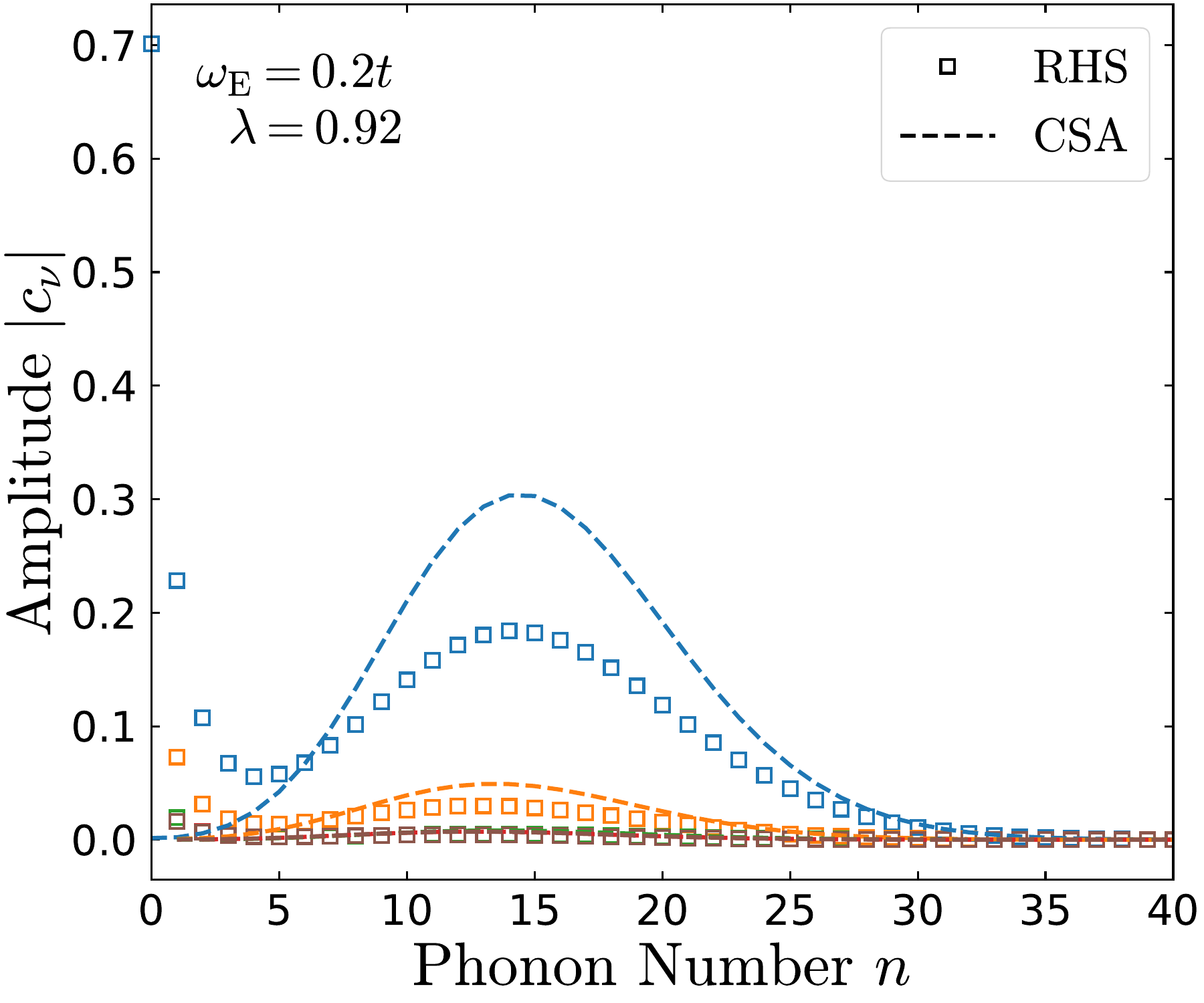}}
    \subfloat[\label{fig_GS_psi_2D_several_lam_c}]{\includegraphics[width=0.32\textwidth]{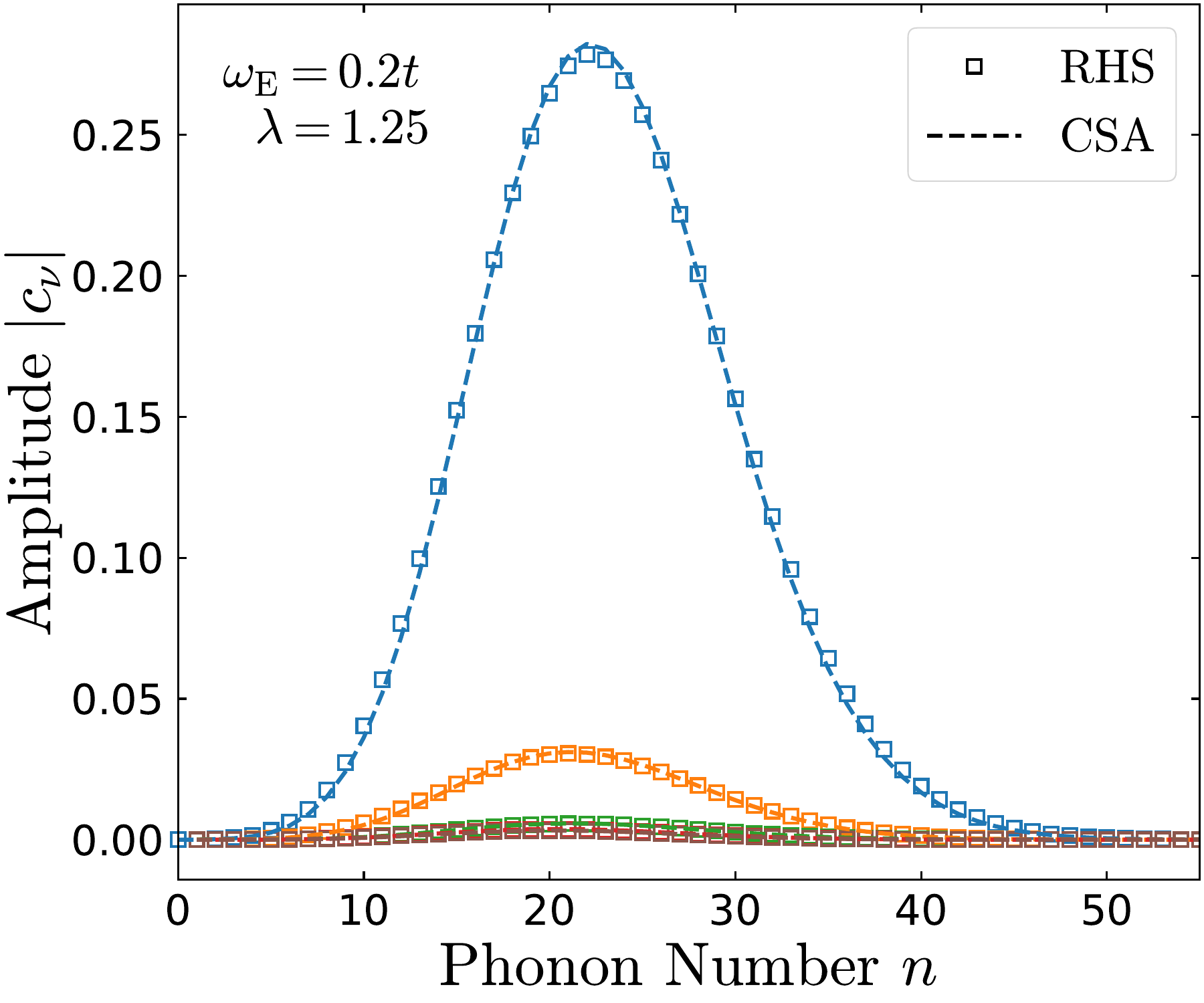}}
    \caption{
    Plot of all components of the RHS and CSA ground-state wavefunctions for $\hw=0.2t$ in 2D. The three panels represent, from left to right, the weak-coupling regime ($\lambda=0.5$), the crossover point ($\lambda=0.92$), and the strong-coupling regime ($\lambda=1.25$). 
    The colour of each curve indicates the family from \cref{fig_1D_states} to which it corresponds. For $n=0$ there is no meaningful distinction between the families, so the entire contribution is assigned to the blue family. In the left and right panels, results from the RHS (square symbols) and CSA (dashed lines) are nearly indistinguishable. Near the crossover, however, the RHS smoothly interpolates between the two types of polaron, while the CSA selects one of the two regimes.}
    \label{fig_GS_psi_2D_several_lam}
\end{figure*}

The CSA is able to faithfully describe two types of polaron, appropriate for the weak- and strong-coupling regimes, respectively. For small values of $\lambda$, the polaron consists of a nearly free electron dressed by a small number of phonons. For large $\lambda$, the polaron contains many phonons, most of which occupy a single site. The plots in \cref{fig_GS_psi_2D_several_lam} show the RHS and CSA predictions for the ground-state wavefunction at three distinct values of $\lambda$: one in the weak-coupling regime, one near the crossover region, and one in the strong-coupling regime. The two methods yield nearly identical results for the weak- and strong-coupling cases, but differ greatly near the crossover (\cref{fig_GS_psi_2D_several_lam_b}). Here, the RHS wavefunction exhibits features of both the weak- and strong-coupling polaron. On the other hand, the CSA must choose one or the other (for the specific value of $\lambda$ shown in the plot, the strong-coupling polaron is favoured). This explains the discontinuous jump in the CSA predictions for the effective mass when moving between the weak- and strong-coupling regimes. Because the CSA is a special case of the RHS approximation, the variational principle dictates that the RHS wavefunction is a better proxy for the exact ground state.

\begin{figure}[ht]
    \centering
    \includegraphics[width=\linewidth]{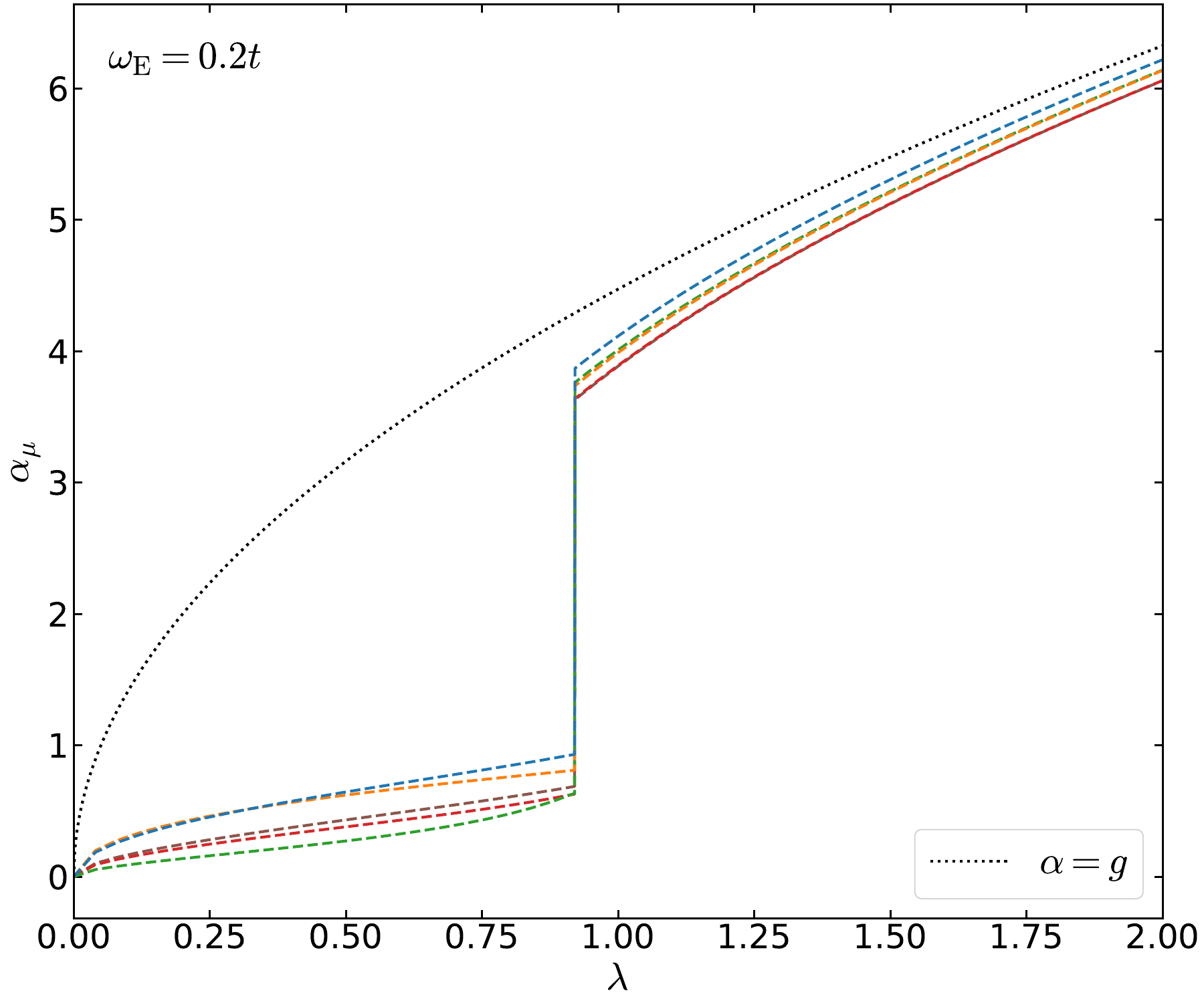}
    \caption{Plot of the variational $\alpha$-values which minimize the CSA energy for $\hw=0.2t$ in 2D. As always, the five families of states are represented by their respective colours as depicted in \cref{fig_1D_states}. Two distinct regimes are evident: a weak-coupling regime with small values of $\alpha_\mu$ and a strong-coupling regime with $\alpha_\mu\sim g$ (the dotted line represents $\alpha=g$). At the critical point $\lambda_{\rm c}=0.92$, there is an abrupt transition between the two regimes in the CSA approximation.}
    \label{fig_alpha_2D}
\end{figure}

To make the CSA crossover behaviour more concrete, we examine in \cref{fig_alpha_2D} the values of $\alpha_\mu$ which minimize the variational energy $\langle \hat{H}\rangle$, as a function of coupling strength. Shown in the plot are the 2D results, but the 1D case is not meaningfully different. For each family $\mu\in \{ {\rm b,\ o,\ g,\ r,\ br} \}$, the mean number of phonons $\langle n_{\rm ph} \rangle_\mu$ in the coherent state is $|\alpha_\mu|^2$. Two distinct types of polaron are readily apparent from the plot. For small $\lambda$, we have $\alpha_\mu<1$, corresponding to $\langle n_{\rm ph} \rangle_\mu < 1$: the electron is dressed by a small number of nearby phonons. Beyond a critical coupling strength $\lambda_{\rm c}=0.92$, we instead have $\alpha_\mu\sim g$, which corresponds to $\langle n_{\rm ph} \rangle_\mu \sim g^2 = 2 d \lambda / (\hw/t)$. For small frequencies, this represents a large number of phonons. This is the heavy polaron which bears little resemblance to a free electron. For coupling strengths near $\lambda_{\rm c}$, the variational energy $\braket{\hat{H}}$ has two local minima, corresponding to the two cases outlined above. For $\lambda<\lambda_{\rm c}$, the weak-coupling polaron is energetically favoured, while the strong-coupling polaron is preferred for $\lambda>\lambda_{\rm c}$. At $\lambda_{\rm c}$, the CSA prediction of a ground-state observable exhibits a phase transition. As we have emphasized, this is not the case for the RHS, nor for the exact solution.

\vspace{-12pt}
\subsubsection{Energy dispersion curves}

\begin{figure*}[t]
    \centering
    \includegraphics[width=\linewidth]{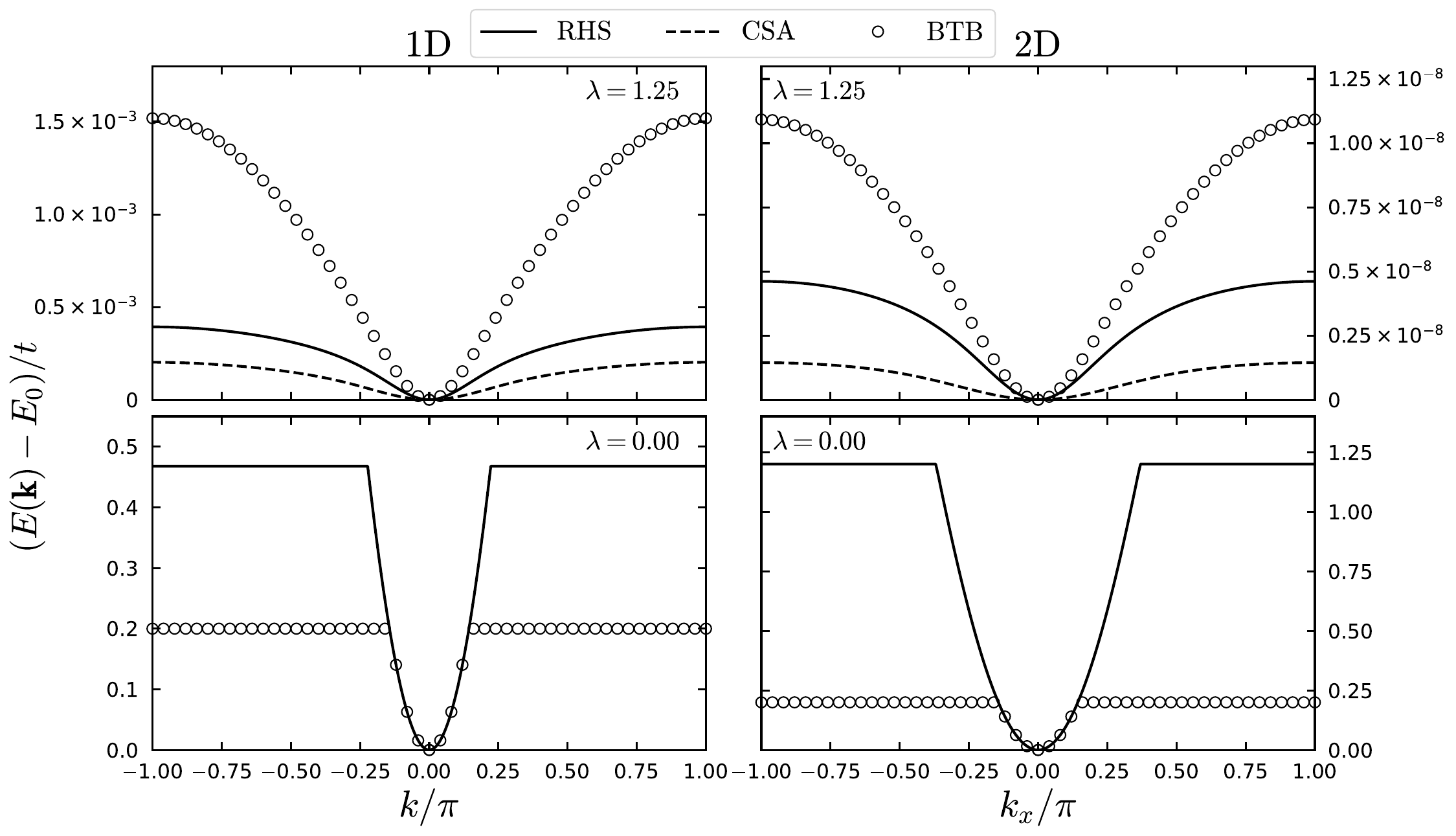}
    \caption{Plots of the energy dispersion $E(\vec{k})$, in 1D (left) and 2D (right), for $\hw=0.2t$. Solid lines represent RHS results while dashed lines represent CSA results and open circles represent exact values calculated using the BTB algorithm. 
    The top set of panels displays values for $\lambda=1.25$, well within the strong-coupling regime, where the bandwidth is significantly suppressed. Here, the approximate methods overestimate the band flatness, though the error in predicted values of $E(\vec{k})$ is small in absolute terms. 
    The bottom set of panels depicts the noninteracting case $\lambda=0$, where the bandwidth is equal to $\we$. CSA results are omitted here, as there is no distinction between some of the coherent-state families in the absence of phonons. Note that physically in the noninteracting case, the exact result is simply the minimum of the approximately parabolic electronic band $E(\vec{k})\simeq k^2/(2m)$ and the flat optical phonon band $E(\vec{k})=\omega_{\rm E}$, hence the nonanalytic behavior at the intersection of the two bands.
    }
    \label{fig_dispersion_curves}
\end{figure*}

Finally, we show select dispersion curves $E(\vec{k})$ for $\hw/t=0.2$ in \cref{fig_dispersion_curves}. The left-hand side of the figure depicts the 1D case, with the 2D results on the right (recall in 2D that we set $k_y=0$ in these plots). The most striking feature in the strong-coupling limit (top row of panels) is that the bandwidth $W = E(\pi) - E(0)$ is extremely small. This is a key feature of the strong-coupling Holstein polaron; in the Lang--Firsov limit~(\ref{LF}), the bandwidth is exponentially suppressed by a factor of $e^{-g^2}$ \cite{langfirsov63}. Much like with the case of the effective mass, our approximative methods here overestimate the flatness of the band, though in absolute terms the energy predictions are quite accurate for all values of $\vec{k}$.

Included in the figure are also the results for the noninteracting case $\lambda=0$ (bottom row of panels). The bandwidth here is $W=\we$, rather than that of a free electron on a lattice for which $W=4t$. To see this, we note that without a coupling term the Hamiltonian $H(\vec{k})$ is exactly solvable, with the lowest-energy state given either by
\begin{equation}
    \ket{\Psi(\vec{k})} = \cre{\vec{k}} \ket{0}\!,\quad E(\vec{k}) = -2t\sum_{i=1}^d \cos k_i \label{small_k_ground_state}
\end{equation}
for small momenta or by
\begin{equation}
    \ket{\Psi(\vec{k})} = \cre{\vec{0}} \creb{\vec{k}} \ket{0}\!,\quad E(\vec{k})= -2dt+\we \label{large_k_ground_state}
\end{equation}
for large momenta. The transition occurs when $\vec{k}$ is such that the two expressions for $E(\vec{k})$ coincide. Importantly, $E(\vec{0})=-2dt$ in the noninteracting case for $\omega_{\rm E}>0$. 
In \cref{small_k_ground_state,large_k_ground_state}, $\cre{\vec{k}}$ ($\creb{\vec{k}}$) creates an electron (a phonon) with momentum $\vec{k}$. We see here that for large $\vec{k}$, and sufficiently small $\we$, the lowest-energy state contains an additional phonon carrying the requisite momentum, allowing the electron to occupy its more energetically favourable zero-momentum state.

We note that the presence of this ``free" phonon means that even the BTB algorithm as presented breaks down when both $\we$ and $\lambda$ are small and $\vec{k}$ is large. The central assumption of BTB is that the phonons in the system remain near the electron, and thus states in which phonons and the electron are greatly separated can safely be discarded. However, to account for the additional phonon in \cref{large_k_ground_state}, this assumption must be relaxed. To achieve exact results for small $\lambda$ and large $\vec{k}$ via the BTB algorithm, we impose an additional truncation $N_{\rm tot}$, discarding states whose total number of phonons in the system exceeds $N_{\rm tot}$. In this way, we are able to include states with a phonon on further-removed lattice sites before reaching computational limits. Compared to the rest of parameter space, the BTB results converge much more slowly here. For similar reasons, the RHS fails quantitatively to capture the lowest-energy state at large $\vec{k}$, as evidenced by the bottom row of \cref{fig_dispersion_curves}. It does, however, qualitatively capture the transition between the two types of ground state. Including more families in the RHS would bridge the gap between the two sets of curves in the figure. The CSA, in which the distinction between families becomes ill-defined as the number of phonons in the coherent states tends to zero, is not well suited to the $\lambda=0$ case, and for this reason we do not include CSA results in the bottom panels of the figure.

\vspace{-8pt}
\section{Conclusion and Outlook}\label{conclusion}

At strong coupling, properties of the Holstein polaron are difficult to compute owing to the large number of phonons present, even in the ground state. We have presented two approximative methods for calculating ground-state polaron properties. These methods are designed specifically to target the strong-coupling regime, and are extremely computationally efficient, making it feasible to apply them to higher-dimensional lattices. Despite their simplicity, they give results which agree reasonably well with more computationally taxing exact methods. 

The first method (CSA) is a variational ansatz, consisting of coherent states of phonons in the vicinity of the electron on the lattice. At strong coupling, it gives very accurate ground-state energies and reasonable approximations to the polaron mass. Ground-state wavefunctions computed using the CSA are similar to the exact results and allow for an intuitive picture of the many-body ground state in terms of families of coherent phonon states surrounding the electron in real space. The second method (RHS) uses exact diagonalization on a restricted subset of the Hilbert space consisting of the individual Fock states contained within the CSA. For strong coupling, the accuracy gained from relaxing the coherent-state constraint is small. For weak and moderate coupling strengths, however, the RHS improves upon the CSA significantly, especially in 2D. The RHS also correctly predicts a smooth crossover in 1D, and a very rapid crossover in 2D, between a weak- and a strong-coupling polaron. 

We should add that the physical nature of this crossover region remains somewhat elusive. It is truly remarkable that in two (and three) dimensions, a typical property like the effective mass, and even the ground-state energy, has such an abrupt change in behaviour at some characteristic coupling strength that is weakly dependent on phonon frequency in the physically relevant regime where $\hw/t\ll1$. This change is abrupt, yet smooth: There is no phase transition, though one has been implied in the past by various simple analytical variational wavefunctions, including the CSA here. Both approximations used here (CSA and RHS) work best at strong coupling and the hope was that they would shed light on the crossover region. They managed to faithfully reproduce accurately the ground-state energy in this region in two dimensions (see \cref{fig_GS_energy_b}), but not the effective mass (see the CSA approximation in particular in \cref{fig_mass_logplot_b}).

Both of these approaches are very readily motivated by the exact results, as \cref{fig_1D_states} shows. However, even in the absence of exact results for other models, provided a Lang-Firsov-like transformation can render the Hamiltonian into a strong-coupling form plus a perturbative hopping part (as in Ref.~\cite{marsiglio95}), strong-coupling perturbation theory will also motivate the relevant sequence of families, as discussed in \cref{coherent}. 

The methods outlined in this paper are simple to use and require very little computational resources. As such, they are promising for obtaining approximate results to more difficult problems. For example, a BTB approach becomes progressively more difficult both for excited states and for the study of nonequilibrium dynamics. However, if accurate results for equilibrium ground-state properties can be established with either the CSA or RHS methods, then both of these topics can be studied, at least in an approximate way. Another natural extension would be to study a two-electron system with an electron-electron repulsion, i.e. with the Hubbard--Holstein Hamiltonian. In this so-called bipolaron problem, already studied in 1D in the regime of intermediate and large $\hw/t$ \cite{trugman2000}, the interaction between polarons is studied in order to determine under which circumstances the electron-phonon interaction leads to the formation of bound pairs. For smaller values of $\hw/t$, it is useful to have access to simple methods which give good approximate descriptions of a polaron with fewer basis states.

In general, we expect both the refinement of the exact method and the application of our approximate techniques to be applicable to the polaron problem for extensions to the Holstein model, for example with a nonlinear $x^2$ interaction \cite{adolphs_2013}, or with slightly longer-range couplings \cite{chandler_2014} and hopping parameters \cite{chandler_2013}. Where our and most other methodologies have difficulty is in models where the phonon ground state, in particular, is non-local, as in the BLF-SSH model.
\cite{BLF_1970,barisic_1972,BLF_1972,SSH_1979,SSH_1980,heeger_1988,chandler_2014_blf-ssh}
Some authors \cite{marchand_2010} have circumvented this difficulty by redefining the SSH model to have an Einstein dispersion relation.

Another promising avenue of exploration would be to treat the Holstein polaron on more complex geometries such as the honeycomb, kagome, or pyrochlore lattices, to name a few that have recently been of interest in condensed matter theory. Obtaining exact solutions to these and other related problems would require significant effort. The CSA, however, can be applied to arbitrary lattices with only minimal tweaking of the procedure described in this work. The RHS, which is extremely computationally efficient compared with other ED methods, could similarly be applied to such problems with modest effort.

\vspace{-8pt}
\begin{acknowledgments}
We acknowledge funding from the Natural Sciences and Engineering Research Council of Canada (NSERC) Discovery Grants RGPIN-2021-02534 and DGECR2021-00043 (CMW and IB) and RGPIN-2022-03295 (CMW and FM), and through a CGRS-M Scholarship (CMW).
\end{acknowledgments}

\appendix
\vspace{-12pt}
\section{CSA calculations}\label[appendix]{appendix_coherent}

We describe here in some detail the procedure for constructing and diagonalizing the coherent-state Hamiltonian matrix. 
Using the adjacency matrix as defined in \cref{A_mat} admits compact expressions for the coherent-state families on an arbitrary lattice. Making explicit the dependence on $\vec{k}$, we express the five variational states as
\begin{align}
    \ket{\psiblue(\vec{k})} &= \frac{1}{\sqrt{N C_{\rm b}(\vec{k})}} \sum_i e^{\i \vec{k} \cdot \vec{R}_i} \cre{i} \ketn{\alphb}{i}\! , \label{psi_blue_app} \\
    \ket{\psiorange(\vec{k})} &= \frac{1}{\sqrt{N C_{\rm o}(\vec{k})}} \sum_{i,j} e^{\i \vec{k} \cdot \vec{R}_i} \bar{A}_{ij}(\vec{k}) \cre{i} \ketn{\alpho}{j}\! , \label{psi_orange_app} \\
    \ket{\psigreen(\vec{k})} &= \frac{1}{\sqrt{N C_{\rm g}(\vec{k})}} \sum_{i,j} e^{\i \vec{k} \cdot \vec{R}_i} A_{ij} \cre{i} \creb{j} \ketn{\alphg}{i}\! , \label{psi_green_app} \\
    \ket{\psired(\vec{k})} &= \frac{1}{\sqrt{N C_{\rm r}(\vec{k})}} \sum_{i,j} e^{\i \vec{k} \cdot \vec{R}_i} \bar{A}_{ij}(\vec{k}) \cre{i} \creb{i} \ketn{\alphr}{j}\! , \label{psi_red_app} \\
    \ket{\psibrown(\vec{k})} &= \frac{1}{\sqrt{N C_{\rm br}(\vec{k})}} \sum_{i,j} e^{\i \vec{k} \cdot \vec{R}_i} (\bar{A}_2)_{ij}(\vec{k}) \cre{i} \ketn{\alphbr}{j}\! , \label{psi_brown_app}
\end{align}
where $i,j=1,\dots,N$, the coefficients $C_{\mu}(\vec{k})$ (with $\mu\in\{\rm b,\ o,\ g,\ r,\ br\}$) are normalization constants, and 
\begin{equation}
    \ket{\alpha^{(j)}}\equiv e^{-\frac{1}{2}|\alpha|^2}e^{\alpha\creb{j}}\ket{0} \label{alpha_def_app}
\end{equation}
denotes a coherent state of phonons on site $j$. Here we have introduced a new notation
\begin{equation}
    \bar{M}_{ij}(\vec{k}) \equiv \meik{\vec{R}_{i}} M_{ij} \eik{\vec{R}_{j}}
\end{equation}
for an arbitrary matrix $M$. In \cref{psi_brown_app}, $A_2$ is the analog of the adjacency matrix for sites separated by two hops. For bipartite Bravais lattices, which we consider in this work, this is given by $A_2 = A^2-q\id$, with $q$ the coordination number of the lattice. Unlike $A$, the matrix elements $(A_2)_{ij}$ are not always $0$ or $1$, but instead count the number of 2-hop paths joining sites $i$ and $j$. The specific choice of phase factors in \cref{psi_blue_app,psi_orange_app,psi_green_app,psi_red_app,psi_brown_app}, as well as the nontrivial weights introduced by the inclusion of $A_2$, were determined empirically from results of the RHS approximation. For the 1D chain with $\vec{k}=0$, the above expressions reduce to the normalized versions of \cref{psi_blue,psi_orange,psi_green,psi_red,psi_brown}.

The coherent states in \cref{psi_blue_app,psi_orange_app,psi_green_app,psi_red_app,psi_brown_app} do not form an orthonormal set; in fact, it can be shown that
\begin{equation}
    \braket{\beta^{(i)}|\alpha^{(j)}} = f(\alpha,\beta) \times \begin{cases} e^{\beta^* \alpha} & \text{if $i=j$,} \\ 1 & \text{if $i\neq j$,} \end{cases} \label{alphabetaoverlap}
\end{equation}
where $f(\alpha,\beta)\equiv \exp\{-\frac{1}{2}(|\alpha|^2+|\beta|^2)\}$. Using this fact, as well as the defining property
\begin{equation}
        \annb{j} \ketn{\alpha}{j} = \alpha \ketn{\alpha}{j}\! ,
\end{equation}
we compute the normalization constants $C_{\mu}(\vec{k})$, the overlap matrix $S_{\mu\nu}(\vec{k}) = \braket{\Psi_\mu(\vec{k})|\Psi_\nu(\vec{k})}$, and the Hamiltonian matrix $H_{\mu\nu}(\vec{k}) = \braket{\Psi_\mu(\vec{k})|\hat{H}|\Psi_\nu(\vec{k})}$. Finally, we perform a Gram--Schmidt orthogonalization and diagonalize the resulting $5\times5$ Hamiltonian matrix to determine the ground-state energy expectation, which is then minimized with respect to the variational parameters $\alpha_\mu$.

In the following, we present the explicit expressions for $C_\mu(\vec{k})$, $S_{\mu\nu}(\vec{k})$, and $H_{\mu\nu}(\vec{k})$. The normalization constants are
\begin{align}
    C_{\rm b}(\vec{k}) &= 1 , \label{norm_blue} \\
    C_{\rm o}(\vec{k}) &= q\Bigl( 1 - e^{-|\alpho|^2} \Bigr) + e^{-|\alpho|^2} \eta_{\vec{k}}^2 , \label{norm_orange} \\
    C_{\rm g}(\vec{k}) &= q , \label{norm_green} \\
    C_{\rm r}(\vec{k}) &= q\Bigl( 1 - e^{-|\alphr|^2} \Bigr) + e^{-|\alphr|^2} \eta_{\vec{k}}^2 , \label{norm_red} \\
    C_{\rm br}(\vec{k}) &= \Bigl( 1 - e^{-|\alphbr|^2} \Bigr) \frac{\Tr(A_2^2)}{N} + e^{-|\alphbr|^2} \bigl(\eta_{\vec{k}}^{(2)}\bigr)^2 , \label{norm_brown} 
\end{align}
where we defined
\begin{align}
    \eta_{\vec{k}}^{\vphantom{(2)}} &\equiv \sum_{j}\bar{A}_{ij}(\vec{k}),\quad \eta_{\vec{k}}^{(2)} \equiv \sum_{j}(\bar{A}_2)_{ij}(\vec{k}).
\end{align} 
For the 1D lattice ($q=2$), we have $\Tr(A_2^2)=2N$ and
\begin{align}
    \eta_{\vec{k}}^{\vphantom{(2)}} &= 2\cos k , \quad
    \eta_{\vec{k}}^{(2)} = 2\cos 2k.
\end{align}
For the 2D lattice ($q=4$), we have $\Tr(A_2^2)=20N$ and
\begin{equation}
\begin{aligned}
    \eta_{\vec{k}}^{\vphantom{(2)}} &= 2\cos k_x + 2\cos k_y , \\
    \eta_{\vec{k}}^{(2)} &= 2\cos(2k_x) + 2\cos(2k_y) + 8\cos k_x \cos k_y .
\end{aligned}
\end{equation}

In the following, we have assumed all $\alpha_\mu\in\mathbb{R}$, because the optimal values are found to be real. The overlap matrix elements are given by ($C_\mu=C_\mu(\vec{k})$)
\begin{align}
    S_{\rm b-o}(\vec{k}) &= \frac{f(\alphb,\alpho)}{\sqrt{C_{\rm o}}} \eta_{\vec{k}}^{\vphantom{(2)}} ,  \\  
    S_{\rm b-r}(\vec{k}) &= \frac{f(\alphb,\alphr)}{\sqrt{C_{\rm r}}} \alphb \eta_{\vec{k}}^{\vphantom{(2)}} ,  \\  
    S_{\rm b-br}(\vec{k}) &= \frac{ f(\alphb,\alphbr)}{\sqrt{C_{\rm br}}} \eta_{\vec{k}}^{(2)} ,  \\ 
    S_{\rm o-g}(\vec{k}) &= \frac{f(\alpho,\alphg)}{\sqrt{C_{\rm o} C_{\rm g}}} \alpho \eta_{\vec{k}}^{\vphantom{(2)}} ,  \\  
    S_{\rm o-br}(\vec{k}) &= \frac{f(\alpho,\alphbr)}{\sqrt{C_{\rm o} C_{\rm br}}} \eta_{\vec{k}}^{\vphantom{(2)}} \eta_{\vec{k}}^{(2)} , \\
    S_{\rm g-r}(\vec{k}) &= \frac{f(\alphg,\alphr)}{\sqrt{C_{\rm g}C_{\rm r}}} \alphg\alphr \eta_{\vec{k}}^{\vphantom{(2)}} , \\  
    S_{\rm b-g}(\vec{k}) &= S_{\rm o-r}(\vec{k}) = S_{\rm g-br}(\vec{k}) = S_{\rm r-br}(\vec{k}) = 0 .
\end{align}
The Hamiltonian matrix elements $h_{\mu\nu}=H_{\mu\nu}(\vec{k})/t$ are
\begin{align}
    h_{\rm b-b} &= -e^{-\alphb^2} \eta_{\vec{k}} + \frac{\hw}{t} \alphb^2 -2g\frac{\hw}{t} \alphb , \\
    h_{\rm b-o} &= -\frac{f(\alphb,\alpho)}{\sqrt{C_{\rm o}}} \Bigl[ q \bigl(e^{\alphb\alpho} - 1\bigr) +\eta_{\vec{k}}^2 + g\frac{\hw}{t} \alphb \eta_{\vec{k}}^{\vphantom{(2)}} \Bigr] , \\ 
    h_{\rm b-g} &= -\frac{f(\alphb,\alphg)}{\sqrt{C_{\rm g}}} \alphb \eta_{\vec{k}}^{\vphantom{(2)}} ,  \\
    h_{\rm b-r} &= \frac{f(\alphb,\alphr)}{\sqrt{C_{\rm r}}} \frac{\hw}{t} \eta_{\vec{k}}^{\vphantom{(2)}} \Bigl[ \alphb -g\alphb^2  - g \Bigr] , \\
    h_{\rm b-br} &= -\frac{f(\alphb,\alphbr)}{\sqrt{C_{\rm br}}} \eta_{\vec{k}}^{(2)} \Bigl[ \eta_{\vec{k}}^{\vphantom{(2)}}  + g\frac{\hw}{t} \alphb \Bigr] , \\
    h_{\rm o-o} &= \frac{1}{C_{\rm o}} \Bigl[ -e^{-\alpho^2} \eta_{\vec{k}}^3 +  \frac{\hw}{t} q\alpho^2 \Bigr] , \\
    h_{\rm o-g} &= \frac{f(\alpho,\alphg)}{\sqrt{C_{\rm o} C_{\rm g}}} \frac{\hw}{t} \alpho \eta_{\vec{k}}^{\vphantom{(2)}} \bigl( 1 - g \alphg \bigr) , \\ 
    h_{\rm o-r} &= -\frac{f(\alpho,\alphr)}{\sqrt{C_{\rm o} C_{\rm r}}} \biggl[ \alpho  q\eta_{\vec{k}}^{\vphantom{(2)}}  +g\frac{\hw}{t} \Bigl( qe^{\alpho\alphr} - q + \eta_{\vec{k}}^2 \Bigr) \biggr] , \\
    h_{\rm o-br} &= -\frac{f(\alpho,\alphbr)}{\sqrt{C_{\rm o} C_{\rm br}}} \biggl[\eta_{\vec{k}}^2 \eta_{\vec{k}}^{(2)}  +\Bigl(e^{\alpho\alphbr} -1 \Bigr) \frac{\Tr(A_2^2)}{N} \biggr] , \\
    h_{\rm g-g} &= \frac{1}{C_{\rm g}} \biggl[ -\alphg^2 e^{-\alphg^2} \eta_{\vec{k}}^{\vphantom{(2)}} + \frac{\hw}{t} q \Bigl( 1 + \alphg^2 - 2g \alphg \Bigr) \biggr] , \\
    h_{\rm g-r} &= \frac{f(\alphg,\alphr)}{\sqrt{C_{\rm g}C_{\rm r}}} \Bigl[ q\bigl(1 - e^{\alphg\alphr} \bigr) - \eta_{\vec{k}}^2 \notag \\
    &\qquad\qquad - \frac{\hw}{t} \eta_{\vec{k}}^{\vphantom{(2)}} \alphr \bigl( g\alphg^2 - 2\alphg + g \bigr) \Bigr] , \\
    h_{\rm g-br} &= - \frac{f(\alphg,\alphbr)}{\sqrt{C_{\rm g}C_{\rm br}}} \frac{\Tr\bigl(\overline{AA}_2(\vec{k}) A\bigr)}{N} \alphbr ,
\end{align}
\begin{align}
    h_{\rm r-r} &= \frac{1}{C_{\rm r}} \biggl[ -\alphr^2 e^{-\alphr^2} \eta_{\vec{k}}^{\vphantom{(2)}} \notag \\
    &\qquad + \frac{\hw}{t} \Bigl(q  + q\alphr^2 + \bigl( \eta_{\vec{k}}^2 - q \bigr) e^{-\alphr^2} \Bigr) \biggr] , \\
    h_{\rm r-br} &= - \frac{f(\alphr,\alphbr)}{\sqrt{C_{\rm r} C_{\rm br}}} g\frac{\hw}{t} \eta_{\vec{k}}^{\vphantom{(2)}} \eta_{\vec{k}}^{(2)} , \\
    h_{\rm br-br} &= \frac{1}{C_{\rm br}} \biggl[ -e^{-\alphbr^2} \eta_{\vec{k}}^{\vphantom{(2)}} \bigl(\eta_{\vec{k}}^{(2)}\bigr)^2 + \frac{\hw}{t} \alphbr^2 \frac{\Tr(A_2^2)}{N} \biggr] .
\end{align}

\vspace{-30pt}
\section{RHS approximation}\label[appendix]{appendix_RHS}

To construct the restricted Hilbert space for arbitrary values of $\vec{k}$, we use the following parametrization of the five families:
\begin{align}
    \ket{\Psi_{\rm b}(\vec{k})} &= \frac{1}{\sqrt{N}} \sum_{i} \eik{\vec{R}_i} \cre{i} \sum_{n\geq 0} d_n^{(\rm b)}\frac{(\creb{i})^n}{\sqrt{n!}}\ket{0}\!, \label{psib_RHS}\\
    \ket{\Psi_{\rm o}(\vec{k})} &= \frac{1}{\sqrt{N}} \sum_{i,j} \eik{\vec{R}_{i}} A_{ij} \cre{i} \sum_{n\geq1} d_{n,j-i}^{(\rm o)}  \frac{(\creb{j})^n}{\sqrt{n!}} \ket{0}\!, \label{psio_RHS} \\
    \ket{\Psi_{\rm g}(\vec{k})} &= \frac{1}{\sqrt{N}} \sum_{i,j} \eik{\vec{R}_i} A_{ij} \cre{i} \sum_{n\geq 1} d_{n,j-i}^{(\rm g)} \creb{j} \frac{(\creb{i})^n}{\sqrt{n!}}\ket{0}\!, \label{psig_RHS} \\
    \ket{\Psi_{\rm r}(\vec{k})} &= \frac{1}{\sqrt{N}} \sum_{i,j} \eik{\vec{R}_i} A_{ij} \cre{i} \sum_{n\geq 2} d_{n,j-i}^{(\rm r)} \creb{i} \frac{(\creb{j})^n}{\sqrt{n!}}\ket{0}\!, \label{psir_RHS} \\
    \ket{\Psi_{\rm br}(\vec{k})} &= \frac{1}{\sqrt{N}} \sum_{i,j} \eik{\vec{R}_i} (A_2)_{ij} \cre{i} \sum_{n\geq 1} d_{n,j-i}^{(\rm br)}\frac{(\creb{j})^n}{\sqrt{n!}}\ket{0}\!, \label{psibr_RHS}
\end{align}
where $d_n^{(\rm b)},\dots,d_{n,j-i}^{(\rm br)}$ are complex weights, to be determined by ED, which depend on $\vec{k}$. Here, the index $j-i$ stands for the site corresponding to $\vec{R}_j-\vec{R}_i$, such that the coefficients $d_{n,j-i}^{(\mu)}$ depend only on the relative separation of sites $i$ and $j$.

The expressions in \cref{psib_RHS,psio_RHS,psig_RHS,psir_RHS,psibr_RHS} simply represent arbitrary linear combinations of states $\ket{\vec{k},\vec{n}}$, as defined in \cref{Bloch_state}, where the form of $\vec{n}$ is chosen to reflect the phonon configuration of the appropriate family. For example, the orange family may be written as
\begin{equation}
    \ket{\Psi_{\rm o}(\vec{k})} = \sum_{j} A_{1j} \sum_{n\geq1} d_{n,j}^{(\rm o)} \ket{\vec{k},(n^{(j)})}\!, \label{psi_orange_RHS}
\end{equation}
where $(n^{(j)})$ denotes a phonon configuration with $n$ phonons on site $j$ and no other phonons in the system. The presence of the adjacency matrix element $A_{1j}$ ensures that site $j$ neighbours the origin, and thus there are only $q$ distinct sets of relevant coefficients $d_{n,j}^{(\rm o)}$.

\begin{figure*}[ht]
    \centering
    \subfloat[\label{fig_mass_linear_a}]{\includegraphics[width=0.5\linewidth]{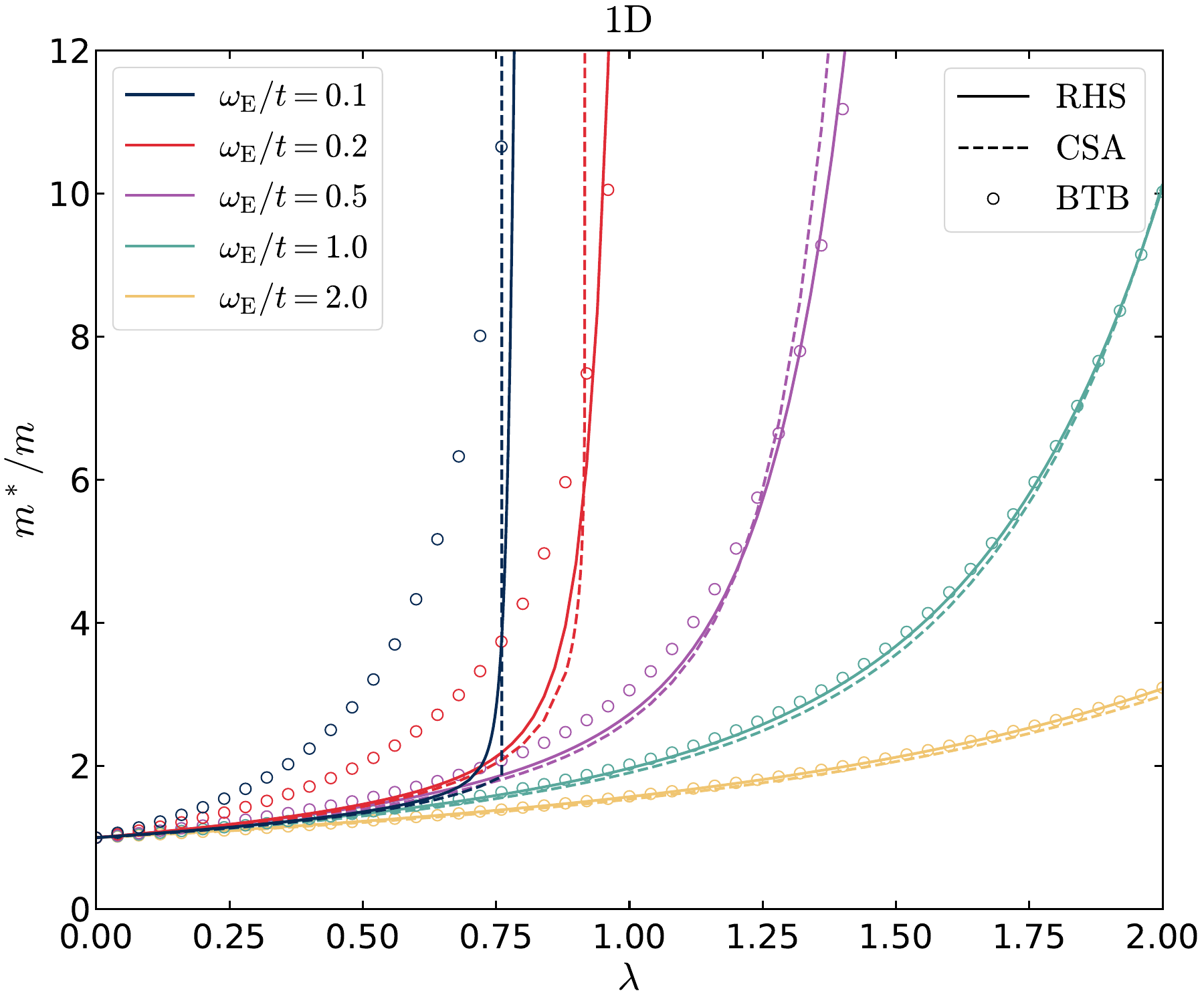}}
    \subfloat[\label{fig_mass_linear_b}]{\includegraphics[width=0.5\linewidth]{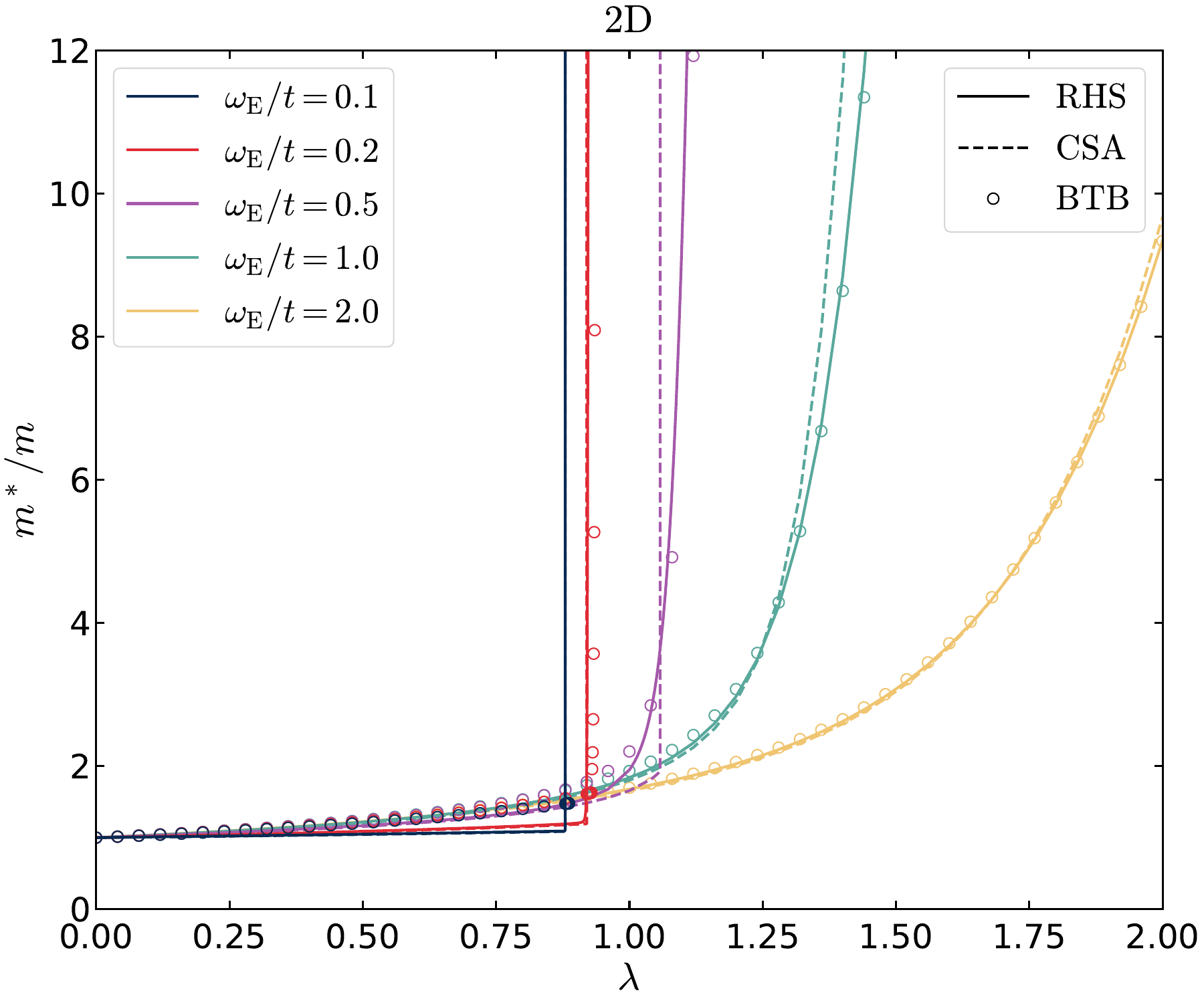}}
    \caption{Effective mass $m^*$ as a function of coupling strength $\lambda$, in 1D (left) and 2D (right), for $\hw/t=0.1,\ 0.2,\ 0.5,\ 1.0,$ and $2.0$. In 1D, the crossover to a large-mass polaron happens gradually and at smaller $\lambda$-values as the frequency is decreased.
    In 2D, there is an extremely abrupt crossover for low frequencies between a weak-coupling regime where the mass is on the order of $m$ and a strong-coupling regime where the mass is orders of magnitude larger. This abrupt crossover is captured well by both the RHS (solid lines) and the CSA (dashed lines). As with the energy, the exact curves are analytic everywhere but nevertheless extremely abrupt at the crossover point.
    }
    \label{fig_mass_linear}
\end{figure*}

Using the definitions from \cref{position_basis,Bloch_state}, we expand \cref{psi_orange_RHS} as
\begin{equation}
\begin{aligned}
    \ket{\Psi_{\rm o}(\vec{k})} &= \frac{1}{\sqrt{N}} \sum_{i,j} \eik{\vec{R}_{i}} A_{1j} \sum_{n\geq1} d_{n,j}^{(\rm o)} \tvec{i} \Bigl( \cre{1} \frac{(\creb{j})^n}{\sqrt{n!}} \ket{0} \Bigr)  \\
    &= \frac{1}{\sqrt{N}} \sum_{i,j} \eik{\vec{R}_{i}} A_{1j} \cre{i} \sum_{n\geq1} d_{n,j}^{(\rm o)} \frac{(\creb{i+j})^n}{\sqrt{n!}} \ket{0}\!, \label{psi_orange_RHS_2}
\end{aligned}
\end{equation}
where again $i+j$ is a notational shorthand for the site located at position $\vec{R}_{i}+\vec{R}_j$. To arrive at the expression in \cref{psio_RHS} from \cref{psi_orange_RHS_2}, we shift the index $j\to j-i$ and note that the translation invariance of the lattice implies that $A_{1,j-i}=A_{ij}$.

An identical procedure can be followed for each of the other families. This yields the expressions in \cref{psib_RHS,psio_RHS,psig_RHS,psir_RHS,psibr_RHS}, or alternatively the parametrizations
\begin{align}
    \ket{\Psi_{\rm g}(\vec{k})} &= \frac{1}{\sqrt{N}} \sum_{i,j} \eik{\vec{R}_i} A_{1j} \cre{i} \sum_{n\geq 1} d_{n,j}^{(\rm g)} \creb{i+j} \frac{(\creb{i})^n}{\sqrt{n!}}\ket{0}\!, \\
    \ket{\Psi_{\rm r}(\vec{k})} &= \frac{1}{\sqrt{N}} \sum_{i,j} \eik{\vec{R}_i} A_{1j} \cre{i} \sum_{n\geq 2} d_{n,j}^{(\rm r)} \creb{i} \frac{(\creb{i+j})^n}{\sqrt{n!}}\ket{0}\!, \\
    \ket{\Psi_{\rm br}(\vec{k})} &= \frac{1}{\sqrt{N}} \sum_{i,j} \eik{\vec{R}_i} (A_2)_{1j} \cre{i} \sum_{n\geq 1} d_{n,j}^{(\rm br)}\frac{(\creb{i+j})^n}{\sqrt{n!}}\ket{0}\!,  \label{psi_brown_RHS_3}
\end{align}
mirroring the form of \cref{psi_orange_RHS_2}.

For appropriate constraints on the weights $d_n^{(\rm b)},\dots,d_{n,j-i}^{(\rm br)}$, each of the expressions in \cref{psib_RHS,psio_RHS,psig_RHS,psir_RHS,psibr_RHS} corresponds to the associated coherent state in the CSA. For example, to recover $\ket{\Psi_{\rm o} (\vec{k})}$ as in \cref{psi_orange_app} from \cref{psio_RHS}, we impose that $d_{n,j-i}^{(\rm o)} \propto \eik{(\vec{R}_{j}-\vec{R}_{i})}(\alpha_{\rm o})^n/\sqrt{n!}$, with $\alpha_{\rm o}$ a variational parameter. Note that in the RHS expressions, the sums over $n$ have different starting indices to avoid including the same basis state in multiple families. To complete the mapping to the CSA expressions, we instead begin all sums at $n=0$. It is this discrepancy in the starting indices that gives rise to the non-orthogonality of the coherent states in the CSA, and thus the need for the normalization factors given in \cref{norm_blue,norm_orange,norm_green,norm_red,norm_brown}.

\section{Polaron effective mass}\label[appendix]{appendix_mass}

The polaron effective mass as plotted in \cref{fig_mass_logplot} grows exponentially quickly with the coupling strength $\lambda$. As such, it can be difficult to discern the weak-coupling region of the plot. We provide in \cref{fig_mass_linear} alternate mass plots with a linear scale on the vertical axis. 

\bibliography{refs_polaron}
\end{document}